\documentclass[a4paper, amsfonts, amssymb, amsmath, reprint, showkeys, nofootinbib]{revtex4-1}
\usepackage[english]{babel}
\usepackage[utf8]{inputenc}
\usepackage{bbm}
\usepackage{yfonts}
\usepackage{xcolor}
\usepackage[colorinlistoftodos, color=green!40, prependcaption]{todonotes}
\usepackage{amsthm}
\usepackage{mathtools}
\usepackage{physics}
\usepackage{xcolor}
\usepackage{graphicx}
\usepackage[left=15mm,right=15mm,top=35mm,columnsep=15pt]{geometry} 
\usepackage{adjustbox}
\usepackage{placeins}
\usepackage[T1]{fontenc}
\usepackage{lipsum}
\usepackage{csquotes}
\usepackage[pdftex, pdftitle={Article}, pdfauthor={Author}]{hyperref} 
\usepackage{qcircuit}
\usepackage{braket}
\usepackage{bbold}
\usepackage{amsmath}

\newcommand{\EqnRef}[1] {Eq.~\eqref{#1}}
\newcommand{\ham}{H}

\begin{document}

\title{Quantum Algorithms for Open Lattice Field Theory}

\author{Jay Hubisz$^{1}$}
\author{Bharath Sambasivam$^{1}$}
\author{Judah Unmuth-Yockey$^{1,3}$}
\affiliation{$^{1}$Department of Physics, Syracuse University, Syracuse, NY 13210}
\affiliation{$^{3}$ Department of Theoretical Physics, Fermi National Accelerator Laboratory, Batavia, IL 60510}


\begin{abstract}
Certain aspects of some unitary quantum systems are well-described by evolution via a non-Hermitian effective Hamiltonian, as in the Wigner-Weisskopf theory for spontaneous decay.  Conversely, any non-Hermitian Hamiltonian evolution can be accommodated in a corresponding unitary system + environment model via a generalization of Wigner-Weisskopf theory.  This demonstrates the physical relevance of novel features such as exceptional points in quantum dynamics, and opens up avenues for studying many body systems in the complex plane of coupling constants. In the case of lattice field theory, sparsity lends these channels the promise of efficient simulation on standardized quantum hardware.  We thus consider quantum operations that correspond to Suzuki-Lee-Trotter approximation of lattice field theories undergoing non-Hermitian time evolution, with potential applicability to studies of spin or gauge models at finite chemical potential, with topological terms, to quantum phase transitions -- a range of models with sign problems.  We develop non-Hermitian quantum circuits and explore their promise on a benchmark, the quantum one-dimensional Ising model with complex longitudinal magnetic field, showing that observables can probe the Lee-Yang edge singularity.  The development of attractors past critical points in the space of complex couplings indicates a potential for study on near-term noisy hardware.
\end{abstract}

\maketitle


\section{Introduction} \label{sec:intro}
   Non-unitary quantum dynamics of lattice field theories are of interest because of their connection to quantum field theories coupled to baths at finite temperature and/or finite chemical potential, or with topological terms. It is also important in the analysis of phase transitions, where the behavior of the Fisher zeros, Lee-Yang zeros, and other features of the partition function at complex values of the parameters give insight into the nature of various thermal and quantum phase transitions (see \cite{CARDY1989275, PhysRevLett.43.805,itzykson_drouffe_1989, UZELAC19801011,vonGehlen:1991zlm,Wei_2014,Itzykson_1986}). Such systems have been studied in the context of quantum computing in \cite{2017,cleve2016efficient,Kliesch_2011,Bertlmann_2006,FESHBACH1962287,Berry_2015,Di_Candia_2015, Motta_2019,Lee_2020,Edvardsson_2019,Kunst_2018,PhysRevA.99.062122,Minganti_2019,PhysRevLett.54.1354}. Furthermore, a class of non-unitary dynamics corresponding to non-Hermitian Hamiltonian dynamics ($\hat{\ham} \neq \hat{\ham}^{\dagger}$) 
   generates a class of models with sign problems, a pressing issue in both condensed matter and high-energy particle physics. Non-Hermitian quantum mechanics has been of great interest in the past couple of decades (see \cite{Bender_2007,Ju_2019}).
   
   To give a specific example: of particular interest are quantum simulations of general spin models.  The lattice $O(N)$ nonlinear sigma models are discretized field theories exhibiting phenomena like confinement, and asymptotic freedom which occur in gauge theories (see \cite{alex2020quantum,PhysRevD.100.054505,Alexandru_2019,Lees_2019,Berry_2014}).  They exhibit quantum phase transitions of various (or infinite) orders, with condensation of topological excitations characterizing the ground state.  At finite chemical potential, less is known about these models.
   
   This is in part due to the fact that classical Monte Carlo studies of such systems are rendered difficult by the sign problem, which occurs because the effective Hamiltonians are non-Hermitian.  There are important known exceptions, where re-parameterization of the theory admits description in terms of new ``dual''
   variables where the partition function is manifestly real and positive, and can be sampled effectively as well as be studied by other analytic methods  (see \cite{Wei_2018,PhysRevD.81.125007,PhysRevLett.106.222001,PhysRevD.92.114508,PhysRevLett.104.112005,WOLFF198592}).
   However, this may not be feasible or possible for all systems (with lattice gauge theory in $D>2$ an important example of physical interest).
   Quantum computing offers the possibility of directly addressing the sign problem, however quantum gates act in a unitary fashion on input quantum states, making it unclear whether non-Hermitian Hamiltonian dynamics can be efficiently simulated.
   
In this work, we describe quantum algorithms for the real-time evolution of a quantum state according to the Schr\"{o}dinger equation with a non-Hermitian Hamiltonian.  This evolution (or an approximation thereof) is accomplished after a Suzuki-Lee-Trotter (SLT) expansion of a unitary time evolution operator.  These unitary time steps are augmented by preparation of, and measurement on, ancillary qubits in order to accomplish the desired non-unitary evolution.

Two algorithms that we present take advantage of the fact that there are unitary system + environment models (a.k.a. unital quantum channels) that are ``close'' to non-Hermitian time evolution.
We construct a Trotterized Lindbladdian with a unitary portion corresponding to the Hermitian component of a non-Hermitian effective Hamiltonian, a non-Hermitian evolution according to the anti-Hermitian component, and additional ``quantum jump'' operations that move the system away from evolution according to the target non-Hermitian Hamiltonian.  The algorithms are an implementation of a damping channel.  
The Liouvillian that corresponds to this nearest unital channel can potentially share physical properties of the non-Hermitian Hamiltonian system of interest, particularly when the non-Hermiticity is small.

Second, we show it is possible to simulate a non-Hermitian Hamiltonian without quantum jumps, but instead each Trotter step is either a step forward or backward in time.
``True'' evolution of the system according to a non-Hermitian Hamiltonian is always achieved, however forward time evolution is not guaranteed.  Ancillary post-selection determines not success or failure, but rather whether a given Trotter step moved the system backwards or forwards in time.  At readout, ancillary measurement outcomes project onto a specific map between simulation time and computational time.
   
In this paper, as a first application, we test these methods on the coarsest ($\mathbf{Z}_2$) discretization of the $O(2)$ nonlinear sigma model: a 1D quantum Ising spin chain in a transverse field.  Non-Hermiticity is introduced via an imaginary longitudinal magnetic field.  The model has been well-studied (see \cite{PhysRevLett.43.805,UZELAC19801011,vonGehlen:1991zlm,itzykson_drouffe_1989}), both on and off the real axis, which makes it an ideal benchmark to test discrete-time, finite-volume quantum algorithms, and to construct and measure observables which probe features of interest such as phase transitions and scaling behavior in the approach to non-unitary critical points.  The Lee-Yang zeros correspond to a non-trivial Jordan-block for the lowest energy pair (by real part) in the eigensystem of the non-Hermitian Hamiltonian.
This model also has a ``true'' sign problem, in that the tensor formulation---or other ``dual-variable'' formulations---of the theory does not eliminate complex phases, and is thus an interesting case study for quantum computation of theories with a sign problem.
   
We show in this paper that these algorithms have the ability to detect the finite size quantum analog of the Lee-Yang edge, where the partition function vanishes due to the effective non-Hermitian Hamiltonian losing an eigenvector.  In quantum simulation, this edge corresponds to a point past which the time evolution develops a fixed point at large times due to the effective ground state energy developing an imaginary part.

 Recently, a noisy intermediate scale quantum (NISQ) algorithm for imaginary time (purely anti-Hermitian) evolution was designed in \cite{Motta_2019}, which is efficient when correlation lengths of the system are small.  In the case of general non-Hermitian hamiltonian evolution, successful SLT evolution can be accomplished by utilizing this ``QITE'' (quantum imaginary time evolution) algorithm for terms in the Hamiltonian with imaginary couplings, and standard algorithms for the unitary part of the time-step. Thus a simple extension of QITE is also applicable for these non-Hermitian Hamiltonians. However, this algorithm would suffer near points of interest such as phase transitions where there is long range order that we wish to emphasize in this paper.  We leave application of QITE to open lattice field theories of this form for future work.

This paper is organized as follows:  In section~\ref{sec:NHHamiltonians}, we give an interpretation of arbitrary non-Hermitian Hamiltonians as open quantum systems, analogous to effective models of heavy particle decay. In section \ref{sec:Koperators}, we talk about modeling non-Hermiticity in a system+environment setting utilizing the formalism of quantum operations. In section \ref{sec:Algs}, we introduce the algorithms for simulating general non-Hermitian Hamiltonians on a quantum computer and write down the corresponding quantum operations. In section \ref{sec:realizations}, we talk about methods to realize these quantum operations. In section \ref{sec:num-tests}, we apply our algorithms to the transverse field Ising model with an imaginary longitudinal field, propose quantum circuits for a small system, and present numerical results of observables and compare with results from exact non-unitary evolution.  We conclude in section \ref{sec:Conclusion}.
   

\section{Non-Hermitian Hamiltonians}\label{sec:NHHamiltonians}

Effective descriptions of quantum many body systems coupled to a bath can, in certain cases, be described or approximated in terms of effective (and often non-Hermitian) Hamiltonians. 
A common simple example of non-Hermiticity elucidated by Feshbach \cite{FESHBACH1962287}, and initially applied to nuclear physics, is that of spontaneous decay of massive resonances, where a small discrete subsystem of at-rest massive particles is weakly coupled to a infinitely sized system of light particles with a continuous range of momenta.  Phase space suppression ensures that information flow is overwhelmingly ``one way'' from the massive system to the light one.

When supplemented with a superselection rule, there is a sense, which we review here, in which tracing out the light particle bath yields evolution via the Schr\"odinger equation with an effective non-Hermitian Hamiltonian.  Dispersive terms in this Hamiltonian account for the decay process. 

Such evolution is not norm preserving (trace preserving in the density matrix formalism), so to make sense of it, we must extend the system, as we now review.
In the case of a simple single particle decay, there is a unitary system + environment model that completes a non-Hermitian Hamiltonian model with $\hat{\ham} = -i \Gamma \mathbbm{1}$ ($\Gamma$ being the width of the particle), and time evolution operator $e^{- \Gamma t}$.  This evolution operator acts on the otherwise trivial 1D Hilbert space spanned by the massive non-interacting particle state, $|M\rangle$.  By supplementing the one dimensional single-particle Hilbert space acted on by $\hat{\ham}$ with a ``de-excited'' vacuum state (the state of no massive particle), the trivial space is promoted to a more physical qubit where measurement yields one of two possible results: ``particle there'' or ``particle gone.''  The operation is the usual amplitude damping channel modeling the loss of energy via some array of possibly unspecified decay processes.  

In the extended Hilbert space, with all decay products traced out, an initial density matrix $\rho_0 = |M \rangle \langle M |$ where the particle is there with certainty, evolves in time $t$ to $\rho_t = e^{-\Gamma t} |M \rangle \langle M | + (1-e^{-\Gamma t}) |0\rangle \langle 0|$, a mixed state with statistics $p_\text{there} = e^{-\Gamma t}$, and $p_\text{gone} = 1-e^{-\Gamma t}$, which obviously preserves the trace condition.

One can also consider more complicated systems with additional massive particles that may interact amongst each other non-trivially, e.g.~with oscillations such as those exhibited by the $K_0$-$\bar{K}_0$ system (see \cite{Bertlmann_2006}).  Such effective Hamiltonians have Hermitian components accommodating oscillation, and anti-Hermitian parts modeling decay.

In essence, we ``make sense'' of these simpler non-Hermitian Hamiltonians via an enlargement of the Hilbert space that accommodates a unital quantum channel.  A sequestered block of a block diagonal density matrix then evolves according to a time evolution operator that is the solution of the Schr\"odinger equation with that non-Hermitian Hamiltonian.

This idea can be generalized to any non-Hermitian Hamiltonian, and we will emphasize that in the case of field theory, with its axiom of local interactions (and the associated sparsity in the corresponding Hamiltonian), such theories may be amenable to simulation on quantum hardware.

To make sense of an arbitrary non-Hermitian Hamiltonian, consider $\hat{\ham}_0= \hat{G}_0 + i \hat{K}_0$, with $\hat{G}_0$ and $\hat{K}_0$ Hermitian.  A sensible (dispersive) model has no eigenvalues with positive imaginary part.  If $\hat{\ham}_0$ has  eigenvalues in the upper half of the complex plane, this can be corrected by subtracting an overall imaginary shift in the Hamiltonian.  

In fact, we shall be more conservative, and shift the vacuum energy to ensure that $-\hat{K}_0$ is a positive semi-definite operator:  $\hat{\ham}_0 \rightarrow \hat{\ham}_0 - i \max ( \text{eigenvalues} (\hat{K}_0))\mathbbm{1}$. This shift creates no change to the relative eigenspectrum of states or their characterization.  It only adds an overall universal decay rate.  The positivity of $-\hat{K}_0$ is required so that small time-steps according to non-Hermitian $\hat{\ham}_0$ can be represented by a unitary system+environment model~\footnote{Some non-Hermitian Hamiltonians have only real eigenvalues (see \cite{Bender_2007}). This subtraction could possibly be unnecessary in these special cases.  Since these systems can be shown to be related to Hermitian but generally non-local Hamiltonians~\cite{Mostafazadeh:2003gz}, they may be intrinsically difficult to simulate without the subtraction we perform.}

Minimally, to accommodate the system decay associated with the shifted Hamiltonian in a unitary quantum channel, the $N$-dimensional Hilbert space acted on by $\hat{H}_0$ must be increased in size by at least one additional basis vector, which we call the ``empty'' state.  We now consider a new Hamiltonian acting on the larger space 
\begin{equation}
    \hat{\ham} = \hat{G} + i \hat{K} \equiv \left( \begin{array}{cc} \hat{\ham}_0 & 0 \\ 0 & 0 \end{array} \right).
\end{equation}
We consider density matrices of the form
\begin{equation}
    \rho = \left( \begin{array}{cc} \rho^\text{Sys}_{N\times N} & 0 \\ 0 & 1-\Tr \rho^\text{Sys}_{N\times N} \end{array}\right).
\end{equation}
Our goal now is to construct a Lindblad formulation~\footnote{A pedagogical treatment of open quantum systems that includes Lindblad evolution and the formalism of quantum operations used here can be found in, for example, Nielsen and Chuang's textbook, Chapter 8~\cite{10.5555/1972505}} of the problem on the new $N+1$ dimensional Hilbert space that does not spoil the superselection rule forbidding superpositions of the system state with the empty state.

We then aim for an evolution of $\rho$ in the $N+1$ dimensional space which follows
\begin{equation}
\label{eq:schrodinger}
    \frac{d\rho}{dt} = - i \left[ \hat{G},\rho \right] - \left\{ \hat{K}, \rho \right\} +\sum_i 2 \hat{L}_i \rho \hat{L}_i^\dagger,
\end{equation}
and where the ``quantum jump'' terms in the sum do not pollute the upper $N\times N$ block of the density matrix. Trace preservation requires that $\hat{K} = \frac{1}{2} \sum_i \hat{L}_i^\dagger \hat{L}_i$.  In $\hat{K}_0$'s eigenbasis, with the imaginary energy shift, we have 
\begin{equation}
    \hat{K}_0 = -\text{diag}(\Gamma_1,\cdots,\Gamma_N),
\end{equation}
where the $\Gamma_i$ are positive decay constants. Our requirements are met with $N$ Lindblad operators given by, 
\begin{equation}
\hat{L}_i =  \left(\begin{array}{cc} 0_{N \times N} & \vec{0}_N \\ \left[\sqrt{-\hat{K}_0}\,\right]_i & 0 \end{array}\right),
\end{equation}
where $\left[\sqrt{-\hat{K}_0}\,\right]_i$ is the $i^{\text{th}}$ row of $\sqrt{-\hat{K}_0}$. For a small discrete time-step, a Trotterized advancement corresponds to an $(N+1)$-element Kraus operator set:
\begin{align}\label{eq:Particle_decay}
    \hat{E}_0 &= \left(\begin{array}{cc} e^{-i \delta t \hat{G}_0} e^{\delta t \hat{K}_0} & 0 \\ 0 & 1 \end{array} \right), \nonumber \\
    \hat{E}_i &= \left( \begin{array}{cc} 0_{N\times N} & \vec{0}_N \\ \left[\sqrt{\mathbb{1}_N-e^{2\delta t \hat{K}_0}}\right]_i & 0 \end{array} \right),
\end{align}
where $\left[\sqrt{\mathbb{1}_N-e^{2\delta t \hat{K}_0}}\right]_i$ is the $i^{\text{th}}$ row of $\sqrt{\mathbb{1}_N-e^{2\delta t \hat{K}_0}}$. The quantum operation described by this set of Kraus operators correctly advances the system up to $\mathcal{O} (\delta t^2)$ terms. Keeping with the particle decay analogy, there are $N$ ``flavors'' of massive particles interacting with one another in the top left block of $\rho$. The Kraus operators $\hat{E}_i$, with $i=1,\cdots,N$ correspond to these different ``flavors'' decaying. 

Over evolution time, the system of interest is decaying into the empty state while undergoing non-Hermitian evolution.  Statistics at late times will tend to be dominated by the empty measurement, with failure of the simulation to yield information about the system of interest.

Novel properties of the non-Hermitian lattice system, as a rule of thumb, would be expected to be manifest at time scales inverse to the anti-Hermitian Hamiltonian term's magnitude.  However, failure probabilities are expected to approach unity at time scales inverse to the system volume multiplied by the anti-Hermitian coupling strengths.  This is the usual price of fitting non-Hermitian evolution into completely positive maps:  naive implementation comes at the cost of growing probability of a ``garbage'' outcome.

For any algorithm that simulates a non-Hermitian Hamiltonian via a trace preserving quantum operation, the probability of success for a single Trotter step 
depends on the way we normalize the desired evolution 
so as to fit it into a trace preserving quantum operation. As we have emphasized, $-\hat{K}_0$ must minimally be positive definite,
and thus probability for a successful time step on an initial density matrix $\rho$ without an undesired quantum jump is bounded:
\begin{equation}
    p_s \leq \Tr \left( \hat{E}_0 \rho \hat{E}^\dagger_0 \right) = \Tr \left( e^{\delta t \hat{K}} \rho e^{\delta t \hat{K}} \right).
    \label{eq:psuc}
\end{equation}

Realistically, $\hat{K}$ will not in general be diagonal in the lattice basis, and we will not know its spectrum.  Instead, for a lattice model, we will have $\hat{K}$ as a sum of locally acting operators, $\hat{K} = \sum \hat{k}_I$ that we implement as individual anti-Hermitian ``gates,'' and for which we do know the spectra.  In a simulation we must enforce positivity of each $-\hat{k}_I$.  This will over-compensate in general, and in a typical simulation the bound in Eq.~(\ref{eq:psuc}) will not be saturated.

We next move on to study the construction of simulations that target non-hermitian Hamiltonian evolution on lattice models.


\section{Modeling anti-Hermiticity}
\label{sec:Koperators}

Our algorithms are general, however in view of simplicity of exposition and also our intention to study lattice field theories with local interactions, we need only put focus on non-Hermitian Hamiltonian terms which are single site or involve interactions between nearest neighbor degrees of freedom.  In this section, we describe a family of Kraus operator decompositions of an arbitrary non-unitary time-step, limits of which correspond to our ``random walk through time'' algorithm with probabilistic time evolution, and damping circuits with straightforward time evolution which minimize failure probability.


\subsection{Single Qubit anti-Hermiticity}
\label{sec:singlequbit}

If the anti-Hermitian part of the Hamiltonian acts only locally on lattice degrees of freedom, the Trotterized transfer matrix can be separated into a unitary part encompassed in $\hat{G}$ and a non-unitary part from $\hat{K}$ which acts only on single system sites.  We focus on finding a \emph{single} qubit quantum operation that approximates a time step according to such an anti-Hermitian Hamiltonian.  It can, in fact, be shown that any Trotterized multi-qubit non-Hermitian evolution can be decomposed into unitaries and single-qubit non-unitary quantum operations, as we explain further in Section~\ref{sec:MultiNonUDecomp}.

Tracing out the entire system with the exception of a single qubit gives a reduced density matrix for the $i$'th lattice site, $\rho_i$.  The portion of the Trotter step that solves \EqnRef{eq:schrodinger} corresponding to anti-Hermitian evolution is
\begin{equation}
    \rho_i \rightarrow e^{\delta t \hat{k}_i} \rho_i e^{\delta t \hat{k}_i},
\end{equation}
where $\hat{k}_i$ here is the portion of $\hat{K}$ acting at site $i$.
This takes a form similar to a single element non-trace-preserving quantum operation, with Kraus operator $\hat{E}^i_0= e^{\delta t \hat{k}_i}$.
As emphasized in Section 2, requiring positivity of $-\hat{k}_i$ guarantees that $\hat{E}_0^{i~^\dagger} \hat{E}_0^i \leq \mathbb{1}$, as required by unitarity.
Without loss of generality, one can always apply a unitary transformation to rotate the single qubit anti-Hermitian evolution to point along the $z$-axis.  We thus consider the specific case $\hat{k}_{i} = \Theta (\hat{\sigma}_z - s \hat{\mathbb{1}})$, 
 so that we have
\begin{equation}
    \hat{E}_{0}^i = \left(\begin{array}{cc}
         e^{(1-s) \delta t \Theta}  & 0 \\
    0     & e^{-(1+s) \delta t \Theta} 
    \end{array}\right),
\end{equation}
where $\Theta$ is a coupling strength, and we require $s \ge 1$.

In the following section, we offer a few simple constructions of Kraus operator sets that can be incorporated as quantum channels in a circuit-based implementation of non-Hermitian quantum dynamics.  Each has its advantages and disadvantages relative to each other.  First, we show that in order to implement two-qubit non-unitary gates, it is sufficient to have the capability to implement one-qubit non-unitary gates.


\subsection{Decomposing Two-Qubit Gates}
\label{sec:MultiNonUDecomp}
The aim in this section is to decompose a general $N$-qubit operator into a single qubit non-unitary operation, and $N$-qubit unitaries. We will first show this for the two-qubit case and then generalize. This is valuable since many lattice field theory interactions are non-linear, and their quantum-computation encodings will necessarily be reduced to single- and two-qubit gates, which may themselves be non-unitary. Being able to reduce two-qubit non-unitary gates, for instance into two-qubit unitary operations, and a single-qubit non-unitary operation allows the implementation of more complicated systems that possess two-qubit gates.

Consider an arbitrary two qubit Trotterized evolution operation of the form
\begin{equation}
    \hat{M}_2=e^{-i\delta t \hat{\ham}_2}, \quad \text{where} \quad \hat{\ham}_2 = \hat{G}_2+i \hat{K}_2.
\end{equation}
Using the Trotter expansion, we can write it as a unitary piece and a non-unitary piece:
\begin{equation}
    \hat{M}_2 \approx \hat{M}_{\text{U}} \hat{M}_{\text{NU}}= e^{-i\,\delta t\, \hat{G}_2}e^{\delta t\, \hat{K}_2}.
\end{equation}
The unitary piece can in principle be implemented on a quantum computer. Now, consider the Pauli decomposition of $\hat{K}_2$. Since $\hat{K}_2$ is Hermitian, the coefficients, $a_{ij}$ of the decomposition will be real
\begin{align}
    \hat{M}_{\text{NU}}&=\exp{\delta t\sum_{i,j}a_{ij}(\hat{\sigma}_i \otimes \hat{\sigma}_j)}\\
    &\approx \prod_{i,j} \hat{M}_{\text{NU}}^{ij} = \prod_{i,j}\exp{\delta t\, a_{ij}(\hat{\sigma}_i \otimes \hat{\sigma}_j)},
\end{align}
where the Trotter expansion has been used in the second equality, and $i,j=0,1,2,3$, with $\hat{\sigma}_0 \equiv \mathbb{1}_2$. Since $\hat{M}_{\text{NU}}^{ij}$ is Hermitian $\forall\,(i,j)$, it admits a spectral decomposition with orthonormal eigenvectors
\begin{equation}
    \hat{M}_{\text{NU}}^{ij}=\hat{U}^{ij}\hat{\Lambda}^{ij} (\hat{U}^{ij})^{\dagger},
\end{equation}
where $\hat{U}^{ij}$ is the unitary matrix comprised of the eigenvectors of $\hat{M}_{\text{NU}}^{ij}$ ordered in the decreasing order of the eigenvalues; $\hat{\Lambda}^{ij}$ is a diagonal matrix with the eigenvalues in decreasing order. One can show that
\begin{equation}\label{eqn:2qubitEigs}
\hat{\Lambda}^{ij}=\begin{cases} 
  \exp{\delta t\,a_{00}\,\mathbb{1}_2} \otimes \mathbb{1}_2 , & (i,j)=(0,0) \\
  \exp{\delta t\,a_{ij}\,\hat{\sigma}_3\,} \otimes \mathbb{1}_2,   & (i,j)\neq (0,0).
 \end{cases}
\end{equation}
Using  this, we can write 
\begin{equation}
    \hat{M}_{\text{NU}}=\hat{\Lambda}^{00}\times\prod_{\substack{(i,j)\neq\\(0,0)}} \hat{U}^{ij}(\exp{\delta t\,a_{ij}\,\hat{\sigma}_3}\otimes \mathbb{1}_2)(\hat{U}^{ij})^{\dagger}.
\end{equation}
This is the required decomposition. The $\hat{U}^{ij}$'s are two qubit unitaries (entanglers), and the operations in the middle are single qubit operations, one of which is always the identity.

This decomposition is easily generalized to non-unitary operations acting on $N$ qubits:
\begin{equation}
    \hat{M}_{\text{NU}}^{\{k_i\}} = \hat{U}^{\{k_i\}}\hat{\Lambda}^{\{k_i\}}(\hat{U}^{\{k_i\}})^{\dagger},
\end{equation}
where the set $\{k_i\}$ is a label for the Pauli basis for a $2^{N}$ dimensional Hilbert space. The (ordered) eigenvalue matrix is then
\begin{equation}
\hat{\Lambda}^{\{k_i\}}=\begin{cases}
  \exp{\delta t\,a_{\{0\}}\,\mathbb{1}_2} \otimes \mathbb{1}_{2^{N}}\, , & k_i=0\,\forall\, i \\
  \exp{\delta t\,a_{\{k_i\}}\,\hat{\sigma}_3} \otimes \mathbb{1}_{2^{N}},   & \exists\, i: k_i\neq 0,
 \end{cases}
\end{equation}
which yields
\begin{multline}
    \hat{M}_{\text{NU}} = \hat{\Lambda}^{\{0\}}\times\\\prod_{\substack{\{k_i\}\neq\\\{0\}}}\hat{U}^{\{k_i\}}\big( \exp{\delta t\,a_{\{k_i\}}\,\hat{\sigma}_3}\otimes\mathbb{1}_{2^N}\big)\big(\hat{U}^{\{k_i\}}\big)^{\dagger}.
\end{multline}
Here, the $\hat{U}^{k_i}$'s are $N$-qubit unitary operators . Even in this general case, the non-unitarity can be moved to be on just a single qubit. Note that these $N$-qubit entanglers need not be efficiently implementable, generally. A block circuit of the two-qubit case is shown in Figure \ref{fig:2qubitDecomp}, where $\Upsilon$ is a unitary implementation of the quantum operations we describe in section~\ref{sec:Algs} to implement single-qubit non-unitary evolution using a single ancillary qubit. $M_Z$ is a measurement of the spin in the $\hat{\sigma}_z$ basis, and $X\equiv \hat{\sigma}_x$.

\begin{figure}[t!]
    \begin{equation*}
    \Qcircuit @C=0.9em @R=0.8em {
    \lstick{\ket{\psi^i_1}\cdots} & \qw & \multigate{1}{\, (U^{ij})^{\dagger} \,}&\qw & \multigate{1}{\, U^{ij} \,}&\qw&\qw&\rstick{\cdots\ket{\psi_1}}  \\
    \lstick{\ket{\psi^i_2}\cdots}  &  \qw & \ghost{\,  (U^{ij})^{\dagger}  \,}&\multigate{1}{\Upsilon(a_{ij})} & \ghost{\,  U^{ij}  \,}&\qw&\qw&\rstick{\cdots\ket{\psi_2}}\\
    \lstick{\ket{0}\cdots}&\qw &\qw &\ghost{\Upsilon(a_{ij})}&\measure{M_Z}&\gate{X^{M_Z}}&\qw&\rstick{\cdots\ket{0}}
    }
\end{equation*}

\begin{equation*}
    \Qcircuit @C=0.9em @R=0.8em {
    \lstick{\ket{\psi_1}} & \qw & \qw & \qw & \multigate{1}{M_\text{U}}& \qw &\qw&\rstick{\ket{\psi^f_1}} \\
    \lstick{\ket{\psi_2}} & \qw & \multigate{1}{\Upsilon(a_{00})} & \qw & \ghost{M_{\text{U}}}& \qw &\qw& \rstick{\ket{\psi^f_2}} \\
    \lstick{\ket{0}} & \qw & \ghost{\Upsilon(a_{00})} &\measure{M_Z}&\gate{X^{M_Z}}&\qw&\qw&\rstick{\ket{0}}
    }
\end{equation*}
\caption{Block circuit of a general Trotterized two qubit operation in terms of single qubit non-unitaries (implemented by $\Upsilon$), and two-qubit unitary entanglers, $U^{ij}$'s.}
\label{fig:2qubitDecomp}
\end{figure}


\section{Explicit quantum channels}
\label{sec:Algs}
In this section, we will elaborate on the specific quantum channels based on the ideas from the previous sections. These are channels obtained by embedding the non-Hermitian system into a bigger, unital channel using ancillary qubits. Then, the undesired evolution can be minimized through post-selection on the measurements of the ancillary qubits. The first two approaches below take the system away from the desired evolution in the eventuality of a wrong measurement on the ancillary qubit. The last one solves this problem at the expense of accumulating Trotter error.

The three channels described here are considered to be implemented uninterrupted for the desired amount of evolution time.  This is how the numerical results in Sec.~\ref{sec:num-tests} were calculated.  However, the channels could just as well be supplemented with tomography of the quantum state to ``checkpoint'' evolution along the way.  In this way all three approaches are able to reproduce the desired evolution up to controllable errors.  


\subsection{System in Decline}
\label{sec:SysEnv}
 The first channel we discuss is modeled on, and inspired by the operation corresponding to particle decay in \cite{Bertlmann_2006}.  Implementation requires that we, at minimum, extend the single qubit system to a qutrit, although for purposes of utilization on standard hardware, we will instead exhibit here a realization via extension by an additional qubit.  We refer to this new qubit as the ``compensatory'' system.  This qubit adds additional states for probability of the system qubit to move into, and so ``compensates'' the ``decay'' experienced by the system qubit.   
To write the Kraus operators, first we will rotate into the $z$-axis and normalize as mentioned in Sec.~\ref{sec:singlequbit}, then let us write down the two matrices
 \begin{align}
     \hat{\Omega}_{1} =
     \begin{pmatrix}
        1 & 0 \\
        0 & \sqrt{1-\gamma} 
    \end{pmatrix},
    \quad
    \hat{\Omega}_{2} =
    \begin{pmatrix}
    0 & \sqrt{\gamma} \\
    0 & 0
    \end{pmatrix}
 \end{align}
 with $\gamma = 1-e^{-4 \delta t \Theta}$.
 Now, a set of two Kraus operators acting on this 4 dimensional Hilbert space is given by
\begin{equation}
\label{eq:decline-kraus}
\hat{E}^{\text{SD}}_{0} =
\begin{pmatrix}
\hat{\Omega}_{1} & 0 \\
0 & \mathbb{1}
\end{pmatrix},
\quad
\hat{E}^{\text{SD}}_{1} =
\begin{pmatrix}
0 & 0 \\
\hat{\Omega}_{2} & 0
\end{pmatrix}.
\end{equation}
To simulate the non-Hermitian system, we impose a super-selection rule, and, on initialization, only consider states which populate the $2\times 2$ blocks lying on the diagonal of the 2-qubit density matrix, with the $2\times 2$ block in the upper left (the 0-state of the compensatory qubit) representing 
the system that will evolve according to the effective non-Hermitian Hamiltonian.

With the Kraus operators in \EqnRef{eq:decline-kraus} this quantum operation is an amplitude-damping channel in which a decay of the system's 1-state, represented by $\hat{E}_0$, is compensated for by population of the 1-state of the compensatory qubit via $\hat{E}_1$.  To begin describing the evolution, we write down a density matrix for the system and compensatory qubit in block-diagonal form, $\rho_{0} = \ket{0}\bra{0} \otimes \rho^{\text{sys}}_{0} + \ket{1}\bra{1} \otimes \rho^{G}_{0}$. The state $\rho^G$ is a ``garbage state'' that contains minimal information about the system that we intend to simulate.  Here, while we use the notation $\rho^{\text{sys}}$ ($\rho^{G}$), these blocks are not themselves individual density matrices. 
Evolving the system an amount of time, $\delta t$, we find
\begin{align}\label{eq:SiDEvol}
\nonumber
    \rho_{0} \rightarrow \rho_{\delta t} &= \ket{0}\bra{0} \otimes \rho_{\delta t}^{\text{sys}} + \ket{1}\bra{1} \otimes \rho^{G}_{\delta t} \\ \nonumber
    &= \ket{0}\bra{0} \otimes \hat{\Omega}_{1} \rho_{0}^{\text{sys}} \hat{\Omega}_{1}^{\dagger}
     \\
     &+ \ket{1}\bra{1} \otimes (\rho_{0}^{G} + \hat{\Omega}_{2}\rho_{0}^{\text{sys}}\hat{\Omega}_{2}^{\dagger}).
\end{align}
On any given time step, the probability, $p_\text{sys}$, for measuring the compensatory qubit in the 0-state after a single step of evolution is given using the maximal case of \EqnRef{eq:psuc} by
\begin{equation}
    p_{\text{sys}} =\Tr\big(\hat{P}_0\rho_{\delta t} \hat{P}_0\big)=\Tr( \hat{\Omega}_{1} \rho_\text{sys} \hat{\Omega}_{1}^{\dagger}), 
\end{equation}
where $\hat{P}_0$ is the projector onto the 0-state of the compensatory qubit. 
In order to actually implement the quantum operation described by Eq.~\eqref{eq:decline-kraus}, a minimal additional third qubit must be introduced for a unitary operation as an ancillary environment, which, when traced out, yields the above Kraus operator set.
With this operation, the upper $2\times 2$ block will evolve, up to normalization, precisely as desired up to computational error in the lab, and finite-time Trotter error. 

This is, however, at the expense of depletion of the system of interest into the garbage state that eventually and inevitably (without tuning) dominates the density matrix, given enough iterations of the non-unitary
portion of the time-step.  Longer system-evolution times are associated with greater expense in performing tomography that yields information about the state of the target system.  

We should note that in some (or many) cases, our damping channel will be overly conservative.  Probabilities of success may be arranged to be higher with more careful construction of the operation.  This is due to the fact that the original non-Hermitian Hamiltonian may have either purely real eigenvalues due to symmetry arguments~\cite{Bender_2007}, or eigenvalues with a positive imaginary part that are overcompensated for in our pursuit of ensuring unital quantum channels to simulate local interactions.  Of course we will not typically know in advance the spectrum of these Hamiltonians (that being the purpose of their simulation), and playing it safe is likely to be best practice.


\subsection{Damping Channels}
\label{sec:algo-damp}
We can also realize the non-Hermitian dynamics without extending the \emph{system} space with a compensatory qubit, but rather use an ancillary qubit to elevate the non-unitary operation to a unitary one.

The simplest way to implement the non-unitary dynamics may be a phase damping quantum channel, with measurement performed on a single ancillary qubit in the computational basis.  
In the framework at the end of Sec.~\ref{sec:NHHamiltonians}, and the beginning of Sec.~\ref{sec:singlequbit}, this case corresponds to minimizing $s$, and having only a single additional Kraus operator complete the set.
Minimizing $s$ provides the single-step optimal probability of obtaining evolution via $\hat{E}^{\text{DC}}_0$ rather than the other Kraus element.  It can be arranged so that a ``0'' outcome of ancillary measurement corresponds to the desired $\hat{E}^{\text{DC}}_0$ evolution, and ``1'' is associated with an undesired $\hat{E}^{\text{DC}}_1$ evolution. 
In this case,
\begin{equation}
\label{eq:damping-kraus}
\hat{E}^{\text{DC}}_0 = \left(\begin{array}{cc}
         1& 0 \\
    0     & \sqrt{1-\gamma} \end{array} \right) 
~~~\text{and}~~~
    \hat{E}^{\text{DC}}_1 = \left(\begin{array}{cc}
         0& 0 \\
    0     & \sqrt{\gamma}
    \end{array}\right),
\end{equation}
with $\gamma = 1 - e^{-4 \Theta \delta t}$.  Implementation of the channel is via a controlled $y$-rotation, $R_y (\phi)$, where the system qubit acts as the control, the ancilla qubit as the target, and $\phi = 2 \sin^{-1} (\sqrt{\gamma})$.
 
For the phase damping circuit, a ``1'' measurement is irrecoverable, destroying any entanglement built up between the $i^{\text{th}}$ qubit and the rest of the system, putting the $i^{\text{th}}$ qubit in the pure state $\rho^\text{1}_i = |1\rangle \langle 1|$.  These 1-measurements correspond to the inevitable quantum jumps associated with a probabilistic
algorithm.  

A low probability to measure the 1-state in the ancillary ensures errors are local and sparse.  Long-range or global properties of the system of interest may survive this approximation to non-Hermitian Hamiltonian dynamics.  We study this numerically in section~\ref{sec:num-tests} by comparing observables in the 1D Ising model at imaginary longitudinal field calculated using the phase damping gate implementation, to those calculated exactly.

We note that one could just as easily consider any right-acting unitary rotation of $\hat{E}^{\text{DC}}_1$, such as $\hat{E}^{\text{DC}}_1 \hat{\sigma}_{x}$ (which in this case would give an amplitude damping channel).


\subsection{Random walk through time}
\label{sec:algo-rwtt}
In contrast to the previous two approaches where a measurement of ``1'' on the ancillary qubit results in a complete loss of entanglement between the qubit and the rest of the system,
in this section, we present an algorithm where an unsuccessful measurement on the ancilla(s) means that the system has evolved in the wrong direction in time by a fraction of the time-step, $\delta t$. Let $\hat{A}_i$ and $\hat{B}_j$ be a set of $N$ and $M$ Kraus operators that define the trace preserving quantum operation given by
\begin{equation}
\label{eq:RWQO}
    \mathcal{R}(\rho)=\sum_{i=1}^{N} \hat{A}_i \rho \hat{A}_i^{\dagger}+\sum_{j=1}^{M} \hat{B}_j \rho \hat{B}_j^{\dagger},
\end{equation}
with
\begin{align}
\label{eq:RWQO_Kraus}
\nonumber
    \hat{A}_i&=\sqrt{\alpha_i}\exp{-i \hat{G} \delta t_i} \exp{\hat{K} \delta t_i},\\
    \hat{B}_j&=\sqrt{\beta_j}\exp{i \hat{G} \delta t'_j}\exp{- \hat{K} \delta t'_j}.
\end{align}
Here, $\hat{A}_i$ does a forward time-step by $\delta t_i$, and $\hat{B}_j$ does a backward time-step by $\delta t'_j$. Note that $\delta t_i, \delta t'_j\leq \delta t$; the equality is when $N=M=1$. 

We demonstrate here a calculation which gives the optimal number of Kraus operators. Under the assumption that $\delta t_i,\, \delta t'_j\ll 1$, the trace-preserving condition ($\sum_i \hat{A}_i^{\dagger} \hat{A}_i+\sum_j \hat{B}_j^{\dagger} \hat{B}_j=1$) gives the following:
\begin{align}
    \nonumber
    \label{eq:RWConditions}
    \sum_i\alpha_i\delta t_i-\sum_j\beta_j\delta t'_j&=0,\\
    \sum_i\alpha_i+\sum_j\beta_j&=1.
\end{align}
The probabilities for the operators $\hat{A}$ and $\hat{B}$ are given by
\begin{align}
\nonumber
    a_i&=\Tr(\hat{A}_i\rho \hat{A}_i^{\dagger})=\alpha_i+2\alpha_i\delta t_i\Tr(\hat{K} \rho)+\order{\delta t_i^2},\\
    b_j&=\Tr(\hat{B}_j\rho \hat{B}_j^{\dagger})=\beta_j-2\beta_j\delta t'_j\Tr(\hat{K} \rho)+\order{\delta t_j'^{\,2}}.
\end{align}
Using these, the average distance in time a single action of the quantum operation in \EqnRef{eq:RWQO} would take the system is
\begin{equation}
    \langle t\rangle=\sum_i a_i\, \delta t_i-\sum_j b_j\, \delta t'_j=0,
\end{equation}
where \EqnRef{eq:RWConditions} was used in the second equality. This indeed is expected since this is a random-walk algorithm. The quantity of interest is the root mean squared distance in time:
\begin{align}
\nonumber
    \sqrt{\langle t^2\rangle}&=\Big(\sum_i a_i\, \delta t_i^2+\sum_j b_j\, \delta t_j'^{\,2}\Big)^{1/2}\\
\nonumber
    &=\Big(\sum_i\alpha_i\delta t_i^2+\sum_j\beta_j\delta t_j'^{\,2}\\
\nonumber
    &\hspace{1.5cm}+2\Tr(\hat{K}\rho)\big(\sum_i\alpha_i\delta t_i^3-\sum_j\beta_j\delta t_j'^{\,3}\big)\Big)^{1/2}\\
\nonumber
    &\leq \delta t\Big(\sum_i\alpha_i+\sum_j\beta_j\\
    &\hspace{1.5cm}+2\Tr(\hat{K}\rho)\big(\sum_i\alpha_i\delta t_i-\sum_j\beta_j\delta t'_j\big)\Big)^{1/2}.
\end{align}
Finally, using \eqref{eq:RWConditions}, we have
\begin{equation}
     \sqrt{\langle t^2\rangle}\leq \delta t.
\end{equation}
This can be maximized using just a single ancillary qubit using the following Kraus operators:
\begin{align}
\label{eq:rw-kraus-ops}
    \hat{E}_{\pm} (\delta t) =\frac{1}{\sqrt{2}}\exp(\mp i \hat{G} \delta t)\exp(\pm \hat{K}\, \delta t),
\end{align}
which has the statistics of a coin that has a state-dependant $\order{\delta t}$ bias. This channel evolves correctly up to $\mathcal{O}(\delta t^2)$.  The drawback of this evolution is that, due to the nature of a random walk, many more Trotter steps are required than $\sim 1/\delta t$ in order to evolve the system the desired amount of time.  This results in an accumulation of Trotter error. Using this method a judicious choice of post-selection must be done- for instance, if we take too many steps in the backwards direction, it would be better to start the simulation over.


\section{Circuit Realization}
\label{sec:realizations}
In this section we give explicit forms of the unitary evolution operators for each of the algorithms discussed in Sec.~\ref{sec:Algs}.  These are expressed in terms of known fundamental gates, and Kraus operators.  We also include a discussion of realizing the non-unitary evolution from a non-Hermitian Hamiltonian using the algorithm of oblivious amplitude amplification by expressing the non-unitary evolution operator as a linear combination of unitaries (see \cite{cleve2016efficient,Berry_2015}).


\subsection{System in Decline realization}\label{sec:algo-SiD}
\begin{figure}[t!]
    \begin{equation}
    \nonumber
\Qcircuit @C=1em @R=.7em {
    \lstick{\ket{0}_a} & \qw & \ctrl{1} & \qw & \qw      & \qw & \gate{R_y(\varphi)} & \qw & \qw & \qw & \ctrl{1} & \qw\\
    \lstick{\ket{0}_c} &  \qw & \qswap   & \qw & \gate{X} & \qw & \ctrl{-1} & \qw & \gate{X} & \qw & \qswap & \qw\\
    \lstick{\ket{s_1}}  & \qw & \qswap \qwx & \qw & \qw &\qw & \ctrl{-1} &\qw & \qw & \qw &\qswap \qwx & \qw
}
\end{equation}
\caption{Gate implementation of the unitary in \EqnRef{eq:decline-unitary} for the System in Decline algorithm for a single qubit Trotterized anti-Hermitian evolution. Here, $\varphi=2\sin^{-1}\sqrt{1-e^{-4\delta t\Theta}}$, and the subscripts $a$ and $c$ denote the ancillary and compensatory qubits, respectively.}
\label{fig:decline-circuit}
\end{figure}
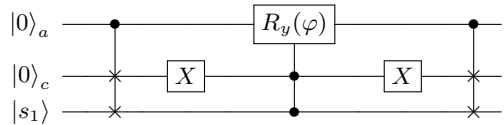

The algorithm described in Sec.~\ref{sec:SysEnv} is best interpreted as a system and ``compensatory'' environment.  The Kraus operators for this algorithm (Eq.~\eqref{eq:decline-kraus}) contain in them how the probability moves between system and environment during evolution.  The actual implementation of these Kraus operators is quite ambiguous, and so one only needs to find a unitary (which is done with an additional ancillary qubit) that applies these Kraus operators correctly.  We found that a unitary that accomplishes this is given by
\begin{multline}
    \label{eq:sys-env-unitary}
    \hat{U}^{\text{SD}}=\frac{1}{2}\Big((\mathbb{1}+ \hat{\sigma}_z)\otimes \hat{E}^{\text{SD}}_0+(\hat{\sigma}_x-i \hat{\sigma}_y)\otimes \hat{E}^{\text{SD}}_1\\ -(\hat{\sigma}_x+i\hat{\sigma}_y)\otimes \hat{E}^{\text{SD}{\dagger}}_1 \\
    + (\mathbb{1}-\hat{\sigma}_z)\otimes(\text{SWAP}) \hat{E}^{\text{SD}}_0 (\text{SWAP})\Big),
\end{multline}
where SWAP is the standard two-qubit swap gate.  Here, the first operator in the tensor product acts on the ancillary qubit, while the second operator acts on the system-compensatory environment qubits. Note that the unitary in \EqnRef{eq:sys-env-unitary} corresponds to the single qubit quantum channel written down in \EqnRef{eq:decline-kraus}, which explicitly gives
\begin{equation}
\label{eq:decline-unitary}
\hat{U}^{\text{SD}}=
    \begin{pmatrix}
        1 & 0 & 0 & 0 & 0 & 0 & 0 & 0\\
        0 & \sqrt{1-\gamma} & 0 & 0 & 0 & 0 & -\sqrt{\gamma} & 0\\
        0 & 0 & 1 & 0 & 0 & 0 & 0 & 0\\
        0 & 0 & 0 & 1 & 0 & 0 & 0 & 0\\
        0 & 0 & 0 & 0 & 1 & 0 & 0 & 0\\
        0 & 0 & 0 & 0 & 0 & 1 & 0 & 0\\
        0 & \sqrt{\gamma} & 0 & 0 & 0 & 0 & \sqrt{1-\gamma} & 0\\
        0 & 0 & 0 & 0 & 0 & 0 & 0 & 1
    \end{pmatrix}.
\end{equation}  
The gate implementation of this unitary is shown in Fig. \ref{fig:decline-circuit}.
Acting with this unitary implements the evolution shown in \EqnRef{eq:SiDEvol}. To see this, consider the combined system-compensatory environment qubit density matrix of the form
\begin{equation}
    \rho=\ket{0}_c\bra{0}_c\otimes\rho^{\text{sys}}+\ket{1}_c\bra{1}_c\otimes\rho^{\text{G}},
\end{equation}
where $\rho^{\text{G}}$ is a garbage state. The total density matrix upon preparing the ancillary qubit in the $1$-state is
\begin{equation}
    \rho_{\text{tot}}=\ket{0}_a\bra{0}_a\otimes\rho,
\end{equation}
where the subscripts $a$ and $c$ correspond to the ancillary and the compensatory environment qubits, respectively. Evolving by $\hat{U}$ yields
\begin{multline}
    \label{eq:decline-unitary-evol}
    \hat{U}^{\text{SD}}\rho_{\text{tot}}\,\hat{U}^{{\text{SD}}\dagger}=\frac{1}{2}\Big((\mathbb{1}+ \hat{\sigma}_z)\otimes \hat{E}^{\text{SD}}_0\rho\hat{E}_0^{{\text{SD}}\dagger}\\+(\hat{\sigma}_x-i \hat{\sigma}_y)\otimes \hat{E}^{\text{SD}}_1\rho\hat{E}_0^{{\text{SD}}\dagger} -(\hat{\sigma}_x+i\hat{\sigma}_y)\otimes \hat{E}^{\text{SD}}_0\rho\hat{E}_1^{{\text{SD}}\dagger} \\
    + (\mathbb{1}-\hat{\sigma}_z)\otimes\hat{E}^{\text{SD}}_1\rho\hat{E}_1^{{\text{SD}}\dagger}\Big).
\end{multline}
Tracing out the ancillary qubit gives the desired evolution in \EqnRef{eq:SiDEvol}.

\subsection{Damping channel realization}
In this section we give an explicit representation for the Kraus operators in the case of the damping circuit, and show their application on an arbitrary quantum state.
In Sec.~\ref{sec:Koperators}, we have considered how to implement non-Hermitian operations on a single qubit, and how to extend it to two qubits.  It turned out that only the technology for single-qubit non-unitary gates is necessary, since non-unitary two-qubit gates can be written in terms of two-qubit unitaries, and a single-qubit non-unitary operation. Then, it reduces to a problem of implementation of the two-qubit unitary entanglers. However, to actually implement the non-unitary operations described in Sec.~\ref{sec:singlequbit}, one must enlarge the space and perform a unitary operation.
There are several equivalent ways to implement the quantum operations described in Sec.~\ref{sec:singlequbit}.  Here, we include an additional ancillary qubit in order to construct a unitary on a larger space.  A unitary that is convenient to utilize is
\begin{equation}
    \label{eq:kraus-unitary}
    \hat{U}^{\text{DC}}=\mathbb{1}\otimes \hat{E}^{\text{DC}}_{0} - i\,\hat{\sigma}_y\otimes \hat{E}^{\text{DC}}_{1},
\end{equation}
where the first operators in the tensor products act on the ancillary qubit, and the second on the system qubit. Acting on the total state $\rho_{\text{tot}}=\left| 0 \right> \left< 0 \right| \otimes \rho$, for example, with $\hat{U}^\text{DC}$ yields the state
\begin{multline}
    \hat{U}^{\text{DC}} \rho_{\text{tot}} \hat{U}^{{\text{DC}}\dagger} = 
    \left| 0 \right> \left< 0 \right| \otimes \hat{E}^{\text{DC}}_0 \rho \hat{E}_0^{{\text{DC}}\dagger}\\ + \left| 1 \right> \left< 1 \right| \otimes \hat{E}^{\text{DC}}_1 \rho \hat{E}_1^{{\text{DC}}\dagger},
\end{multline}
which upon post-selection on the ancillary qubit yields the desired evolution according to $\hat{E}_0$. 
Using the representations from \EqnRef{eq:damping-kraus}, and \EqnRef{eq:kraus-unitary}, we can write the unitary corresponding to the damping gate as,
\begin{align}
    \hat{U}^{\text{DC}} = 
    \begin{pmatrix}
        1 & 0 & 0 & 0 \\
        0 & \sqrt{1-\gamma} & 0 & \sqrt{\gamma} \\
        0 & 0 & 1 & 0 \\
        0 & -\sqrt{\gamma} & 0 & \sqrt{1-\gamma}
    \end{pmatrix}.
    \label{Rsingle}
\end{align}
Looking at the above matrix, this is nothing more than a controlled $y$-rotation, with the system qubit as the control.

Now, consider an arbitrary single-qubit operator, $\hat{M}_1$.  $\hat{M}_1$ admits a singular value decomposition,
\begin{align}
    \hat{M}_1 = \hat{U} \hat{\lambda} \hat{V}^{\dagger}.
\end{align}
where both $\hat{V}$ and $\hat{U}$ are unitary, and $\hat{\lambda}$ is a diagonal matrix with positive entries and let us assume, without loss of generality, that the entries are sorted from largest to smallest: $\lambda_{1} \geq \lambda_2$.  If we normalize by $\lambda_1$, the matrix $\hat{\lambda}$ is identical to $\hat{E}^{\text{DC}}_{0}$ from Eq.~\eqref{eq:damping-kraus}, with $\lambda_2 / \lambda_1 \equiv \sqrt{1-\gamma}$.
Then to implement the matrix $\hat{M}_1$, one applies the matrix $(\mathbbm{1}\otimes \hat{U)} \hat{U}^{\text{DC}} (\mathbbm{1}\otimes \hat{V}^{\dagger})$ to the state $\ket{0}\ket{\psi}$.  In this way, any single qubit matrix which is invertible can be implemented.

By considering the state using the single-site reduced density matrix, $\rho_i$, in standard $(r,\theta,\varphi)$ Bloch-ball 
coordinates, the probability of successful implementation, $p_s$, is the probability associated with obtaining $0$ on a measurement of the ancillary qubit, as given by \EqnRef{eq:psuc}
\begin{equation}
    p_\text{s} = \Tr \left( \hat{E}_0^{\text{DC}} \rho_i \hat{E}_0^{{\text{DC}}\dagger} \right) = 1 - \frac{\gamma}{2} (1-r \cos \theta),
\end{equation}
which is identical to the probability of success in the System in Decline algorithm, since $\hat{E}^{\text{DC}}_0$ and $\hat{\Omega}_1$ are the same. We note that this is independent of implementation/completion of $\hat{E}^{\text{DC}}_0$ to a complete measurement protocol, and bounded below:  $p_s \ge 1-\gamma \approx 1-4 \delta t \Theta$.  Using this lower bound, we can compute the probability of success after $N_{t}$ applications of the operation.  If we set $\delta t = t / N_{t}$,
\begin{align}
    p_{s}^{N_{t}} \geq (1 - 4 \Theta t/N_{t})^{N_{t}},
\end{align}
which, for large $N_{t}$, approaches $e^{-4 \Theta t}$.  Then, the probability of success after $N_{t}$ applications is bounded from below by $e^{-4 \Theta t}$.

This bound for success identifies a line of constant reliability for the circuit.  We see that if $\Theta t = c$ with $c$ small the circuit has a relatively high probability of success after many uses.  That is, the circuit works well along the line starting with $\Theta$ large which is run for small times, and ending with $\Theta$ small but run for long times.  Notice the success of the circuit is independent of the underlying physics of the model.  The probability of success is the same regardless of correlation length, or phase symmetries, and is merely controlled by the quantity $\Theta t$.

The extension of the above to two-qubit gates is a straightforward generalization of the single-qubit case, however unnecessary, as we have already seen in Sec.~\ref{sec:MultiNonUDecomp}, one only needs the technology for single-qubit non-unitary gates to implement two-qubit non-unitary gates when the gate is a by-product of Trotterization. 


\subsection{Random walk realization}
\label{sec:rw-real}
The steps in Sec.~\ref{sec:algo-rwtt} show that by including a single additional qubit, we achieve the best possible evolution using the random walk method. For a general Hamiltonian, $\hat{\ham}=\hat{G}+i \hat{K}$ we would like to be able to identify a Hamiltonian which corresponds to the enlarged unitary evolution.  This would allow for a more straightforward gate interpretation.  In this section we derive it explicitly.

Consider the full, enlarged, unitary, time-evolution operator, separated into the original unitary evolution from the Hermitian part of the Hamiltonian, $\hat{G}$, and the non-unitary evolution from the non-Hermitian part, $\hat{K}$,
\begin{align}
    \hat{U}^{\text{RW}} &\approx \hat{W} \hat{T},
\end{align}
where $\hat{W}$, $\hat{T}$, are the time-evolution operators for $\hat{G}$, and $\hat{K}$, respectively.  Explicitly,
\begin{align}
    \hat{W} = \frac{1}{2}(\mathbbm{1} + \hat{\sigma}_{z}) \otimes e^{-i \hat{G} \delta t} + \frac{1}{2}(\mathbbm{1} - \hat{\sigma}_{z}) \otimes e^{i \hat{G} \delta t}
\end{align}
and
\begin{align}
    \hat{T} = \frac{1}{\sqrt{2}}\mathbbm{1} \otimes e^{\hat{K} \delta t} - \frac{1}{\sqrt{2}} i \hat{\sigma}_{y} \otimes e^{- \hat{K} \delta t}.
\end{align}
First we consider $\hat{T}$, and expand to order $\mathcal{O}(\delta t)$, and collect terms with $\hat{G}$, and $\hat{K}$,
\begin{align}
    \hat{T} &\simeq \frac{1}{\sqrt{2}}\left[(\mathbbm{1} - i \hat{\sigma}_{y}) \otimes \mathbbm{1} + \delta t (\mathbbm{1} + i \hat{\sigma}_{y}) \otimes \hat{K} \right] \\ \nonumber
    &= \frac{1}{\sqrt{2}}((\mathbbm{1} - i \hat{\sigma}_{y}) \otimes \mathbbm{1})[
    \mathbbm{1}\otimes\mathbbm{1} + \frac{\delta t}{2} (\mathbbm{1}+i \hat{\sigma}_{y})^{2} \otimes \hat{K}] \\ \nonumber
    &= \frac{1}{\sqrt{2}}((\mathbbm{1} - i \hat{\sigma}_{y})\otimes \mathbbm{1})[
    \mathbbm{1}\otimes\mathbbm{1} +  i \delta t \hat{\sigma}_{y} \otimes \hat{K}] \\ \nonumber
    &\simeq \frac{1}{\sqrt{2}} ((\mathbbm{1}-i \hat{\sigma}_{y})\otimes \mathbbm{1}) e^{i \delta t \hat{\sigma}_{y} \otimes \hat{K}} + \mathcal{O}(\delta t^2),
\end{align}
where we have worked up to linear order in $\delta t$ and restored the corrections explicitly in the last step.
This is a unitary time-evolution operator, where the non-Hermitian part of the original Hamiltonian is now coupled to the ancillary qubit through a $\hat{\sigma}_{y}$ interaction.

Second, the unitary part, $\hat{W}$, is similar.  Expanding to linear order in $\delta t$ and collecting similar terms,
\begin{align}
\nonumber
    \hat{W} &= \frac{1}{2}(\mathbbm{1} + \hat{\sigma}_{z}) \otimes e^{-i \hat{G} \delta t} + \frac{1}{2}(\mathbbm{1} - \hat{\sigma}_{z}) \otimes e^{i \hat{G} \delta t} \\ \nonumber
    &\simeq (\mathbbm{1} \otimes \mathbbm{1})
     -i \delta t (\hat{\sigma}_{z} \otimes \hat{G}) \\
     &\simeq e^{-i \delta t \hat{\sigma}_{z} \otimes \hat{G}} + \mathcal{O}(\delta t^2).
\end{align}
Then, the Hermitian part of $\hat{\ham}$ can be simulated with an expanded Hamiltonian where the Hermitian part is coupled to the ancillary qubit through a $\hat{\sigma}_{z}$ interaction.  These forms allow Hamiltonian simulation regardless of the form of $\hat{G}$ and $\hat{K}$.  If a qubit formulation for $\hat{G}$ and $\hat{K}$ can be found, those qubit interaction terms can be expanded to include a $\hat{\sigma}_{z}$, $\hat{\sigma}_{y}$ interaction, respectively, in order to simulate the full Hamiltonian using the random-walk method.


\section{The Transverse Ising model with an imaginary longitudinal field}
\label{sec:num-tests}
    \label{sec:ising}
To put these above algorithms into practice, and test the realizations proposed above, we consider a simple lattice model whose Hamiltonian is non-Hermitian: the one dimensional quantum Ising model with a real transverse field and a purely imaginary longitudinal field~\footnote{This can also be understood, in discrete time evolution, as the non-trivial part of the transfer matrix for the 2D Euclidean classical Ising model with an imaginary external field.},
\begin{align}
\label{eq:ising}
    \hat{\ham}_{\text{Ising}} = - \sum_{\langle i j \rangle} \hat{\sigma}^{z}_{i} \hat{\sigma}^{z}_{j} -  \frac{h_{x}}{\lambda}\sum_{i} \hat{\sigma}^{x}_{i} + i \frac{\Theta}{\lambda} \sum_{i} \hat{\sigma}_{i}^{z}.
\end{align}
We re-scale all the couplings by the nearest-neighbor coupling, and will omit it in the following, \emph{i.e.} set $\lambda = 1$.

Brute force classical Monte-Carlo simulations on the discretized imaginary time (Euclidean) partition function (the 2D classical Ising model with an imaginary external field) exhibits a sign problem and disastrous numerical convergence due to the imaginary field. Due to its relative simplicity, however, the model admits study with analytic methods, and much is known about the structure of the phase diagram (see \cite{PhysRevLett.43.805,UZELAC19801011,vonGehlen:1991zlm,itzykson_drouffe_1989}).  The model is thus an ideal benchmark scenario for testing real-time evolution algorithms.

With the longitudinal field set to zero and at large volume, the model exhibits a second order quantum phase transition at $h_x = 1$, where the system switches from a disordered phase to an ordered (magnetized) one.  In a dual description, the transition occurs due to the condensation of topological ``kink'' excitations.  The critical point is associated with a conformal field theory where ungapped kink-antikink bound pairs mediate long-range order, and the entropy of the ground state diverges along with the correlation length in the thermodynamic limit. 

The non-Hermitian extension of the model with imaginary longitudinal field offers another viewpoint on the phase transition.  For $h_x > 1$, there is an ``exceptional line,'' defined by $\pm\Theta_c (h_x)$ along which the ground state (defined here as the eigenvector(s) with smallest real part), merges with the first excited state. This is not a typical degeneracy or level crossing, but rather one at which the non-Hermitian Hamiltonian can be diagonalized only up to a Jordan-block form.  Thus, the system ``loses'' an eigenvector along this line.  This corresponds to a zero in the generating functional for correlation functions (the vacuum to vacuum transition amplitude). 

\begin{figure}[t!]
    \centering
    \includegraphics[width=\linewidth]{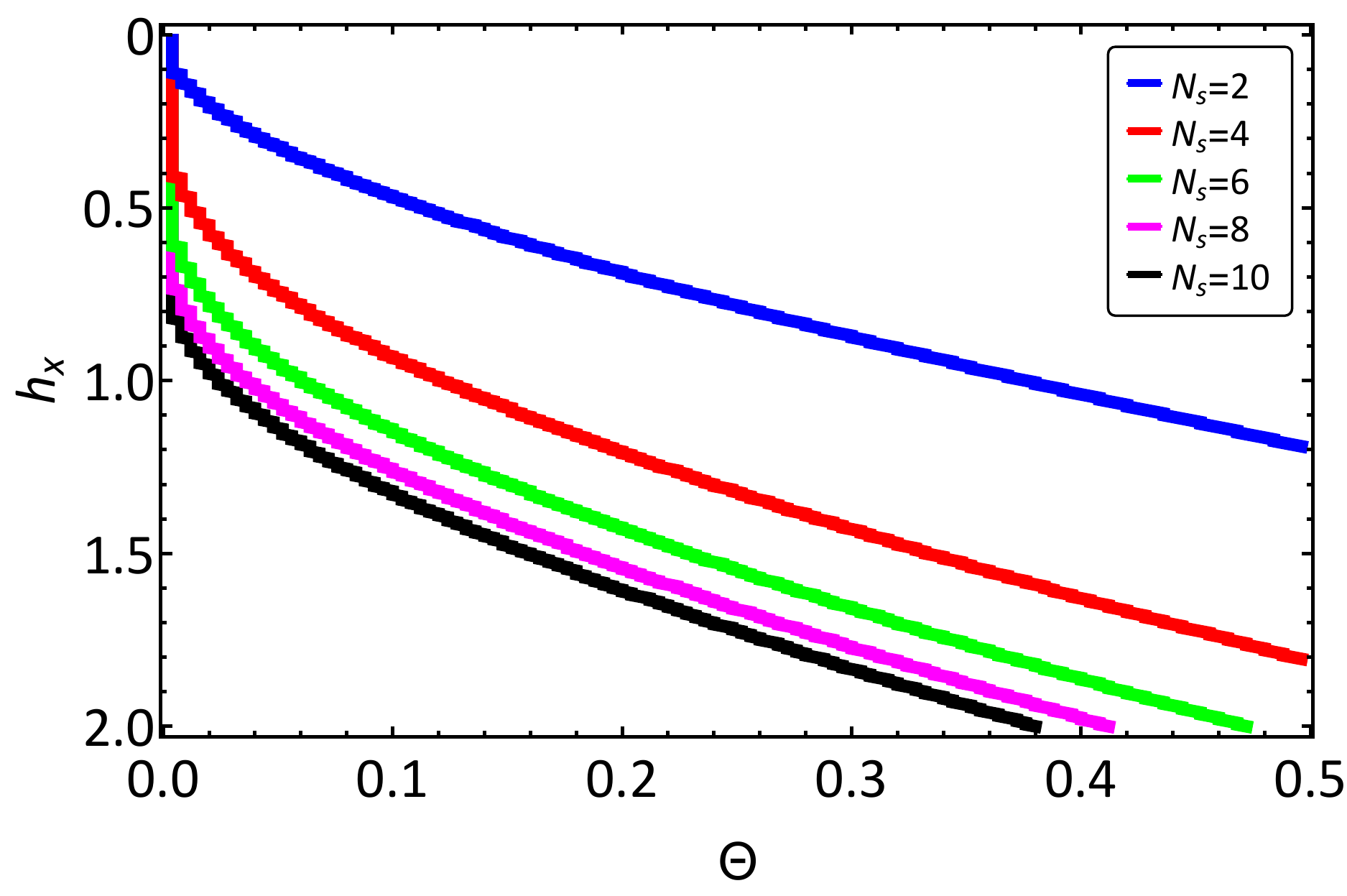}
    \caption{The exceptional line, $(\theta,h_x)_c$ for different system sizes. In the infinite volume limit, the exceptional line deviates from the $\Theta=0$ line at $h_x=1$}
    \label{fig:LYEvN}
\end{figure}

For $|\Theta| > \Theta_c (h_x)$, the ground state is degenerate in its real part, but is separated into a pair of states with energies that are complex conjugate paired.  At large volume, the exceptional line converges toward the $\Theta = 0$ line at $h_x=1$, the location of the quantum phase transition for the original Hermitian system. In Fig. \ref{fig:LYEvN}, we show the exceptional line for different system sizes. This is the quantum analog of the Lee-Yang edge---zeros which lie densely on a circle  (as a function of $e^{i \beta \Theta}$) in the statistical partition function in the large volume limit.  Above the critical $h_x$, the zeros lie outside of a wedge enclosing the real axis.  As $h_x$ is reduced towards the critical point, the wedge closes, and the zeros cluster densely in the immediate vicinity of the real axis.  In the thermodynamic limit, the partition function in the $h_x$-$\Theta$ plane develops a branch-singularity along the $h_x$ axis.

The zeros in the classical partition function map to the line of exceptional points associated with the non-Hermitian Hamiltonian.  They also correspond to a non-unitary critical point in the model; a CFT with central charge $c=-22/5$ (see \cite{vonGehlen:1991zlm,Itzykson_1986}).  The merger (and annihilation) of the exceptional points coincides with the usual 2D Ising CFT.

Despite its non-unitarity, aspects of the complexified Ising model described above admit quantum simulation on a unitary machine.  The techniques described in Sec.~\ref{sec:Algs} can be applied, and aspects of the structure of the non-unitary lattice theory can be probed.  In part, the success of the methods can be traced to the non-unitary features of the model---in particular the effective ground state energies becoming complex.

\begin{figure}[t!]
    \centering
    \includegraphics[width=\linewidth]{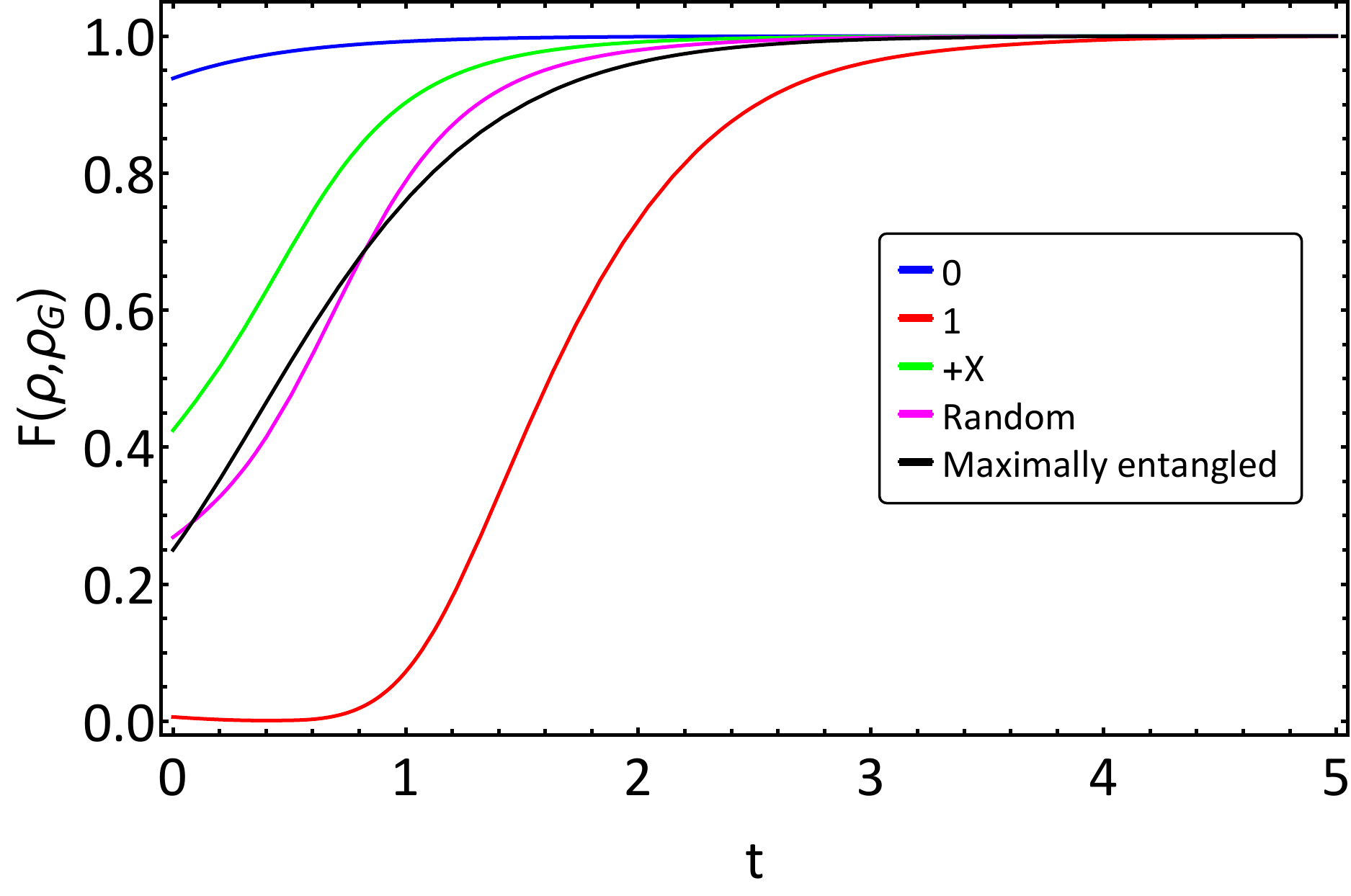}
    \caption{Fidelity between the state of the system and the ground state ($\Theta=h_x=0.5$) for five initial configurations of the system- all in the 0-state, all in the 1-state, all in the +X state, every qubit in a random pure state, and the maximally entangled state.}
    \label{fig:InitState}
\end{figure}

The imaginary part of the ground state energy past the exceptional line leads to the domination (from arbitrary initial configuration) of the ground state in the long-time limit of system evolution. We give an example of this with $4$ system sites in Fig.~\ref{fig:InitState}, where we plot the fidelity, as defined in \EqnRef{eq:Fidelity} of the state of the system and the ground state as a function of time for different initial configurations. This is insensitive to quantum noise that may be associated with the algorithm itself (as in Sec.~\ref{sec:algo-damp}), or, if sufficiently quiet, from environmental noise as well. This makes our algorithms viable for ground state preparation past the Lee-Yang edge. The authors are hopeful that this will create new inroads not only for studying non-unitary models, but also for learning about their real-value limits.  In other words, exploration of the behavior of complex structure of the partition function can herald typical real-space quantum phase transitions in fully unitary theories.

In the next subsections, we describe explicit application of the algorithms in Secs.~\ref{sec:Algs},~\ref{sec:realizations} to the imaginary longitudinal field Ising model in \EqnRef{eq:ising}, constructing gate sequence protocols that are generalizable in principle to arbitrary volume. We simulate Trotter evolution and show how observables such as R\'{e}nyi entropies can distinguish the exceptional line in the $h_x$-$\Theta$ plane. The R\'{e}nyi entropies are good observables because they are sensitive to the interesting physics of the Lee-Yang edge. Moreover, it is not hard to measure them experimentally (see \cite{islam2015measuring,PhysRevA.98.052334,PhysRevB.96.195136}).
The measurements rely on interfering identical copies of the system with each other (through applications of the SWAP gate).  In this way the $n^{\text{th}}$-order R\'{e}nyi entropy is probed through measurements of the parities of sub-systems of one of the copies.  The second-order R\'{e}nyi entropy is the simplest, requiring only two copies.

\subsection{Quantum circuit - System in Decline}
\label{sec:qc-sys-decline}
The anti-Hermitian part of the Ising model with an imaginary longitudinal field described by the Hamiltonian in \EqnRef{eq:ising} can be implemented straightforwardly using the unitary in \EqnRef{eq:sys-env-unitary}. The Kraus operators for this model are
\begin{equation}
\label{eq:SiD-Ising-Kraus}
\hat{E}^{\text{SD}}_{0} =
\begin{pmatrix}
1 & 0 & 0 & 0 \\
0 & \sqrt{1-\gamma} & 0 & 0 \\
0 & 0 & 1 & 0 \\
0 & 0 & 0 & 1
\end{pmatrix},
\quad
\hat{E}^{\text{SD}}_{1} =
\begin{pmatrix}
0 & 0 & 0 & 0 \\
0 & 0 & 0 & 0 \\
0 & \sqrt{\gamma} & 0 & 0 \\
0 & 0 & 0 & 0 
\end{pmatrix},
\end{equation}
with $\gamma=1-e^{-4\delta t \Theta}$. The circuit for a single time-step for this model is shown in Fig.~\ref{fig:SiD-circuit} for two system qubits with one ancillary qubit and one compensatory qubit for each of them. 

Because the probability of success for this circuit is identical to that of the damping channel circuit we leave the numerical results to that section (see Sec.~\ref{sec:phase-damp-circuit}).  Moreover, while the physical set-up is perhaps more intuitive, the gate structure is more complicated than in the damping circuit case, and so for ease of numerical simulation we only study the damping channel circuit.


\subsection{Quantum circuit - Phase damping circuit implementation}
\label{sec:phase-damp-circuit}
The Ising model with a real transverse field, and a purely imaginary longitudinal field can be implemented almost immediately using the phase damping circuit discussed in Sec.~\ref{sec:algo-damp}.  In this case, the Kraus operator, $\hat{E}_{0}$ corresponds to
\begin{align}
    \hat{E}^{\text{SD}}_{0} = 
    \begin{pmatrix}
    1 & 0 \\
    0 & e^{-2 \delta t \Theta}
    \end{pmatrix}.
\end{align}
The actual circuit for a single time step is shown in Fig.~\ref{fig:qo-circuit} for four physical spins, supplemented by four auxiliary qubits which are used to implement the non-unitary gates.

\begin{figure*}[t]
    \begin{equation}
    \nonumber
\Qcircuit @C=1em @R=.7em {
    \lstick{\ket{0}_a} & \qw & \qw & \qw & \qw & \qw & \ctrl{1} & \qw & \qw      & \qw & \gate{R_y(\varphi)} & \qw & \qw & \qw & \ctrl{1} & \qw\\
    \lstick{\ket{0}_c} & \qw & \qw & \qw & \qw & \qw & \qswap   & \qw & \gate{X} & \qw & \ctrl{-1} & \qw & \gate{X} & \qw & \qswap & \qw\\
    \lstick{\ket{s_1}}  & \qw & \ctrl{3} & \qw & \ctrl{3} & \gate{R_x(\delta t h_x)} & \qswap \qwx & \qw & \qw &\qw & \ctrl{-1} &\qw & \qw & \qw &\qswap \qwx & \qw\\
    \lstick{\ket{0}_a} & \qw & \qw & \qw & \qw & \qw & \ctrl{1} & \qw & \qw      & \qw & \gate{R_y(\varphi)} & \qw & \qw & \qw & \ctrl{1} & \qw\\
    \lstick{\ket{0}_c} & \qw & \qw & \qw & \qw & \qw & \qswap   & \qw & \gate{X} & \qw & \ctrl{-1} & \qw & \gate{X} & \qw & \qswap & \qw\\
    \lstick{\ket{s_2}}  & \qw & \targ & \gate{R_z(\delta t)} & \targ & \gate{R_x(\delta t h_x)} & \qswap \qwx & \qw & \qw & \qw & \ctrl{-1} & \qw & \qw & \qw &\qswap \qwx & \qw\\
}
\end{equation}
\caption{A single Trotter step for a two spin Ising system with an imaginary longitudinal field using the System in Decline channel (see Sec.~\ref{sec:algo-SiD}).  Here, $\varphi = 2 \sin^{-1}{\sqrt{1-e^{-4 \delta t \Theta}}}$. The quantum channel is implemented using one ancillary qubit (denoted $a$) and one compensatory qubit (denoted $c$). The compensatory qubits are projected onto the 0-state after the desired amount of evolution.}
\label{fig:SiD-circuit}
\end{figure*}

\begin{figure*}[t]
    \begin{equation}
    \nonumber
\Qcircuit @C=1em @R=.7em {
   \lstick{\ket{0}} & \qw & \qw & \qw & \qw & \gate{R_{y}(\varphi)}  & \measure{M_{Z}} & \gate{X^{M_Z}} & \qw & \qw & \qw \\
   \lstick{\ket{s_{1}}} & \ctrl{2} & \qw & \ctrl{2} & \gate{R_{x}(\delta t h_{x})} & \ctrl{-1} &
   \qw & \qw & \qw & \qw & \qw \\
   \lstick{\ket{0}} & \qw & \qw & \qw & \qw & \qw & \qw & \qw & \gate{R_{y}(\varphi)} & \measure{M_{Z}} & \gate{X^{M_Z}} \\
   \lstick{\ket{s_{2}}} & \targ & \gate{R_{z}(\delta t)} & \targ & \ctrl{2} & \qw & \ctrl{2} & \gate{R_{x}(\delta t h_{x})} & \ctrl{-1} & \qw & \qw \\
   \lstick{\ket{0}} & \qw & \qw & \qw & \qw & \qw & \qw & \qw & \gate{R_{y}(\varphi)} & \measure{M_{Z}} & \gate{X^{M_Z}}  \\
   \lstick{\ket{s_{3}}} & \ctrl{2} & \qw & \ctrl{2} & \targ & \gate{R_{z}(\delta t)} & \targ & \gate{R_{x}(\delta t h_{x})} & \ctrl{-1} & \qw & \qw \\
   \lstick{\ket{0}} & \qw & \qw & \qw & \qw & \gate{R_{y}(\varphi)} & \measure{M_Z} & \gate{X^{M_Z}} & \qw & \qw & \qw  \\
   \lstick{\ket{s_{4}}} & \targ & \gate{R_{z}(\delta t)} & \targ & \gate{R_{x}(\delta t h_{x})} & \ctrl{-1} & \qw & \qw & \qw & \qw & \qw
}
\end{equation}
\caption{A single Trotter step for a four spin Ising system with an imaginary longitudinal field.  Here $\varphi = 2 \sin^{-1}{\sqrt{1-e^{-4 \delta t \Theta}}}$.  In this circuit a measurement is defined as $M_{z}$ which will return either zero or one.  Subsequently the state is flipped depending on the result.}
\label{fig:qo-circuit}
\end{figure*}
\begin{figure}[t!]
    \centering
    \includegraphics[width=8.6cm]{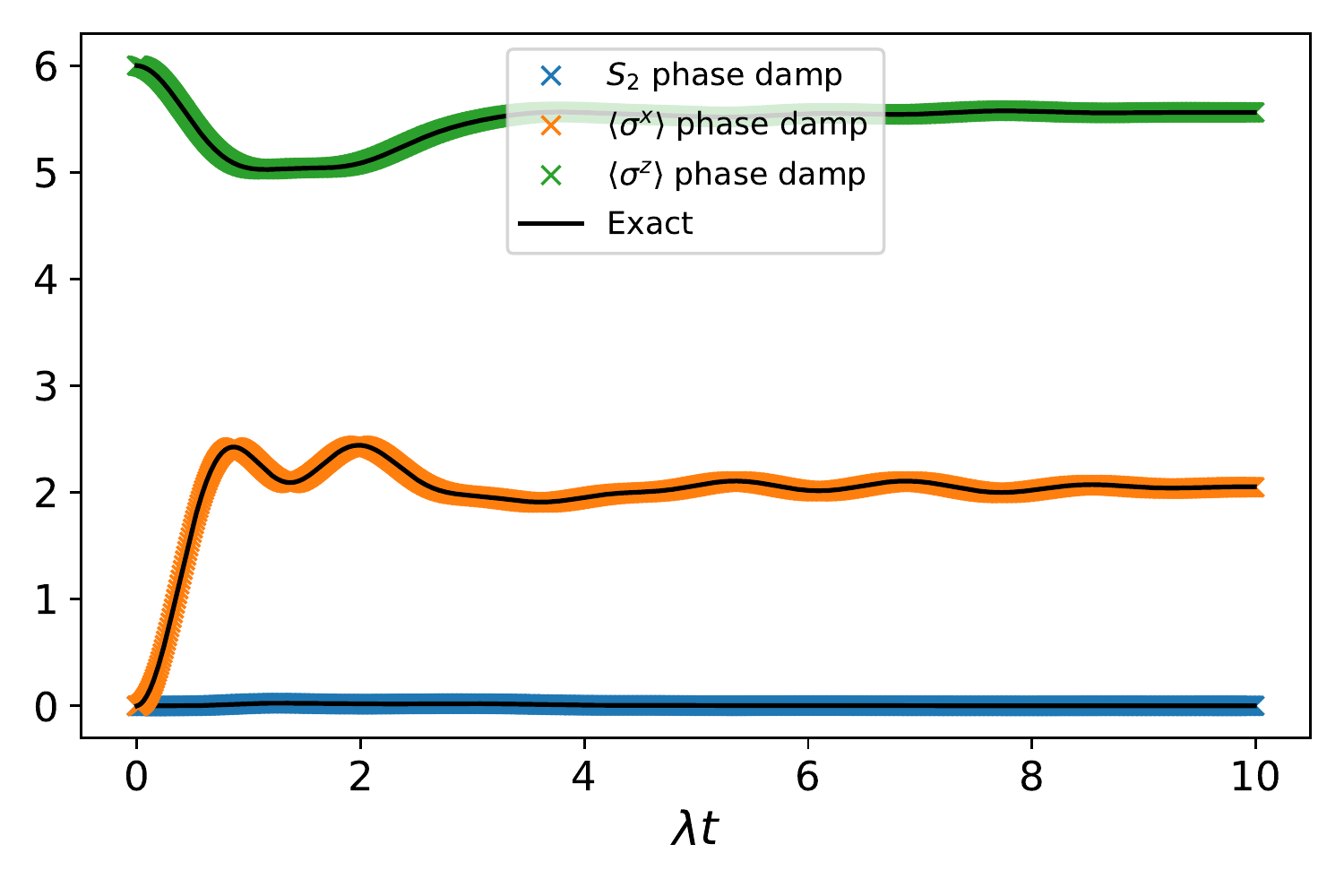}
    \caption{A comparison between the exact non-unitary evolution of six spins using the Ising Hamiltonian from Eq.~\eqref{eq:ising}, and that using a Trotterized circuit like the one shown in Fig.~\ref{fig:qo-circuit}.  The time is in units of the nearest-neighbor coupling, with a Trotter step size $\delta t = 0.01$.  Here $\langle \sigma^x \rangle$ is the expectation value of the $\sigma^{x}$ term in Eq.~\eqref{eq:ising}.  Similarly for $\langle \sigma^{z} \rangle$.  $S_{2}$ is the second-order R\'{e}nyi entropy using a bipartite split.  Here $h_{x} = 0.5$ and $\Theta = 0.1$ placing this data in the ordered phase.}
    \label{fig:dis-obs-compare}
\end{figure}
\begin{figure}[t!]
    \centering
    \includegraphics[width=8.6cm]{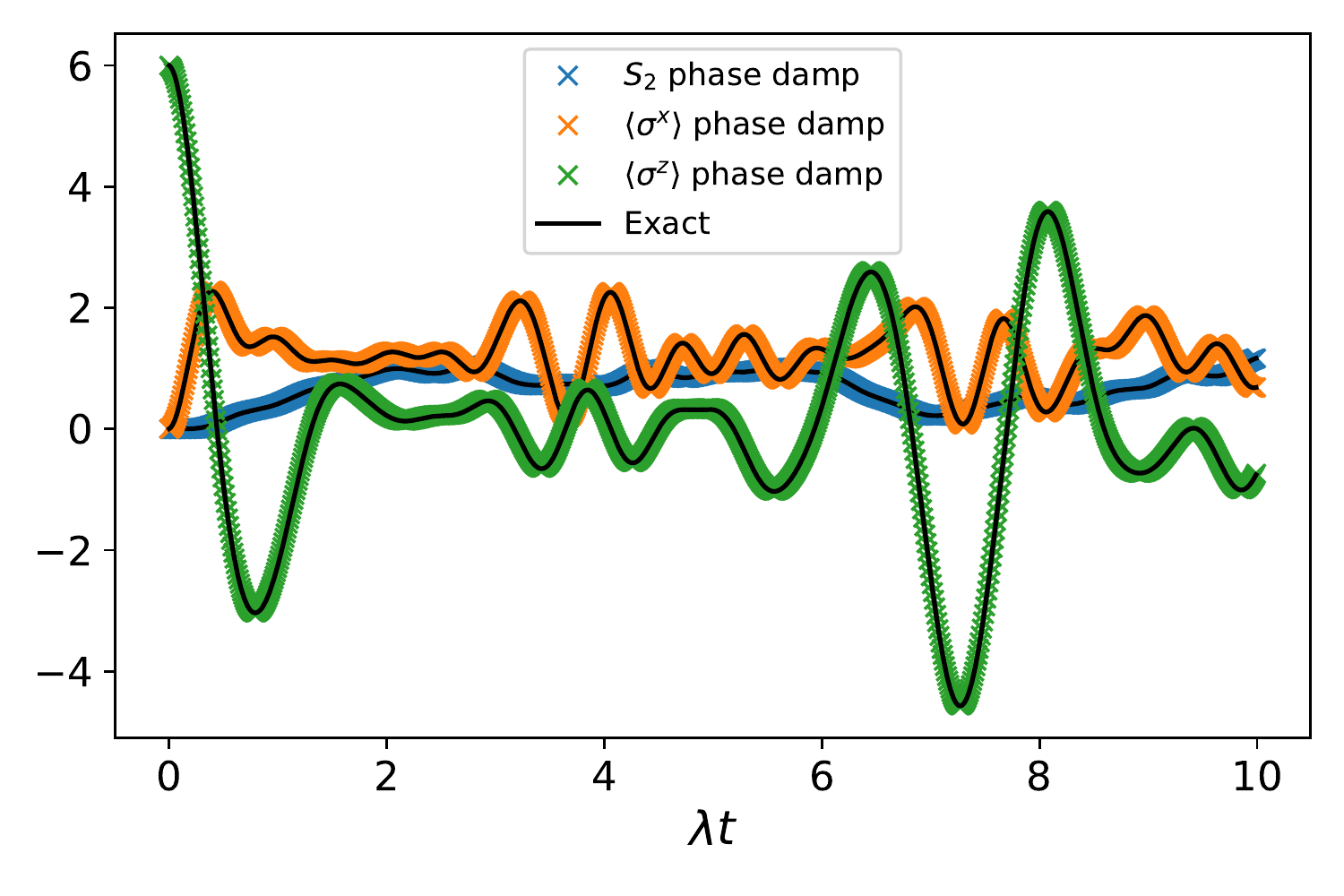}
    \caption{A comparison between the exact non-unitary evolution of six spins using the Ising Hamiltonian from Eq.~\eqref{eq:ising}, and that using a Trotterized circuit like the one shown in Fig.~\ref{fig:qo-circuit}.  The time is in units of the nearest-neighbor coupling, with a Trotter step size $\delta t = 0.01$.  Here $\langle \sigma^x \rangle$ is the expectation value of the $\sigma^{x}$ term in Eq.~\eqref{eq:ising}.  Similarly for $\langle \sigma^{z} \rangle$.  $S_{2}$ is the second-order R\'{e}nyi entropy using a bipartite split.  Here $h_{x} = 2$, and $\Theta = 0.1$ placing this data in the disordered phase.}
    \label{fig:ord-obs-compare}
\end{figure}

A comparison between the method and the exact evolution for six spins can be seen in Fig.~\ref{fig:dis-obs-compare} in the ordered phase.
In the absence of errors it is clear that the circuit, and the method, reproduce the original dynamics almost perfectly.  Another example, this time in the disordered phase, can be seen in Fig.~\ref{fig:ord-obs-compare}.

\begin{figure*}[t]
\begin{minipage}[b]{.325\linewidth}
\centering\includegraphics[width=\textwidth]{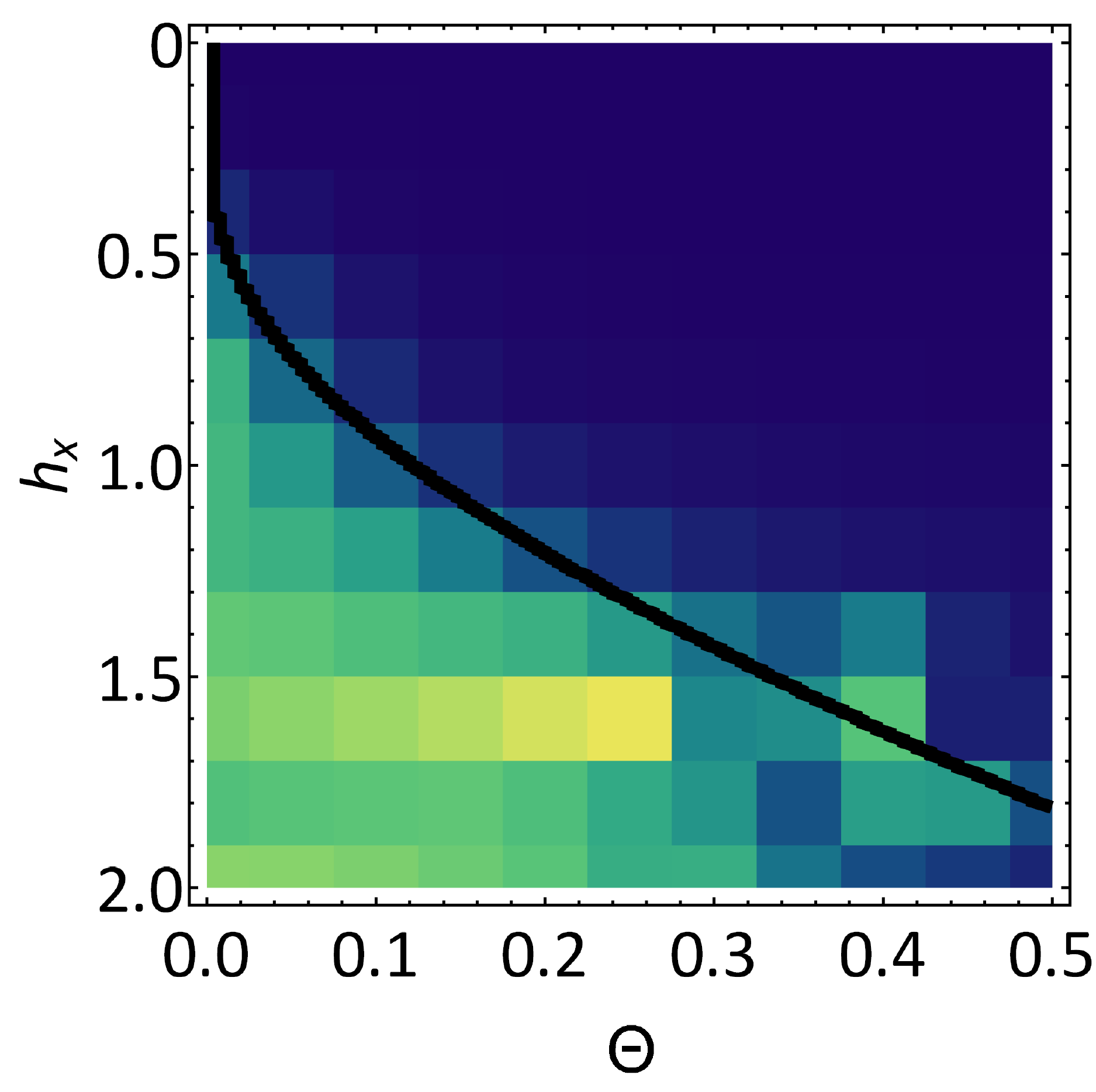}
\label{fig:S2_Damping}
\end{minipage}%
\hfill
\begin{minipage}[b]{.325\linewidth}
\centering\includegraphics[width=\textwidth]{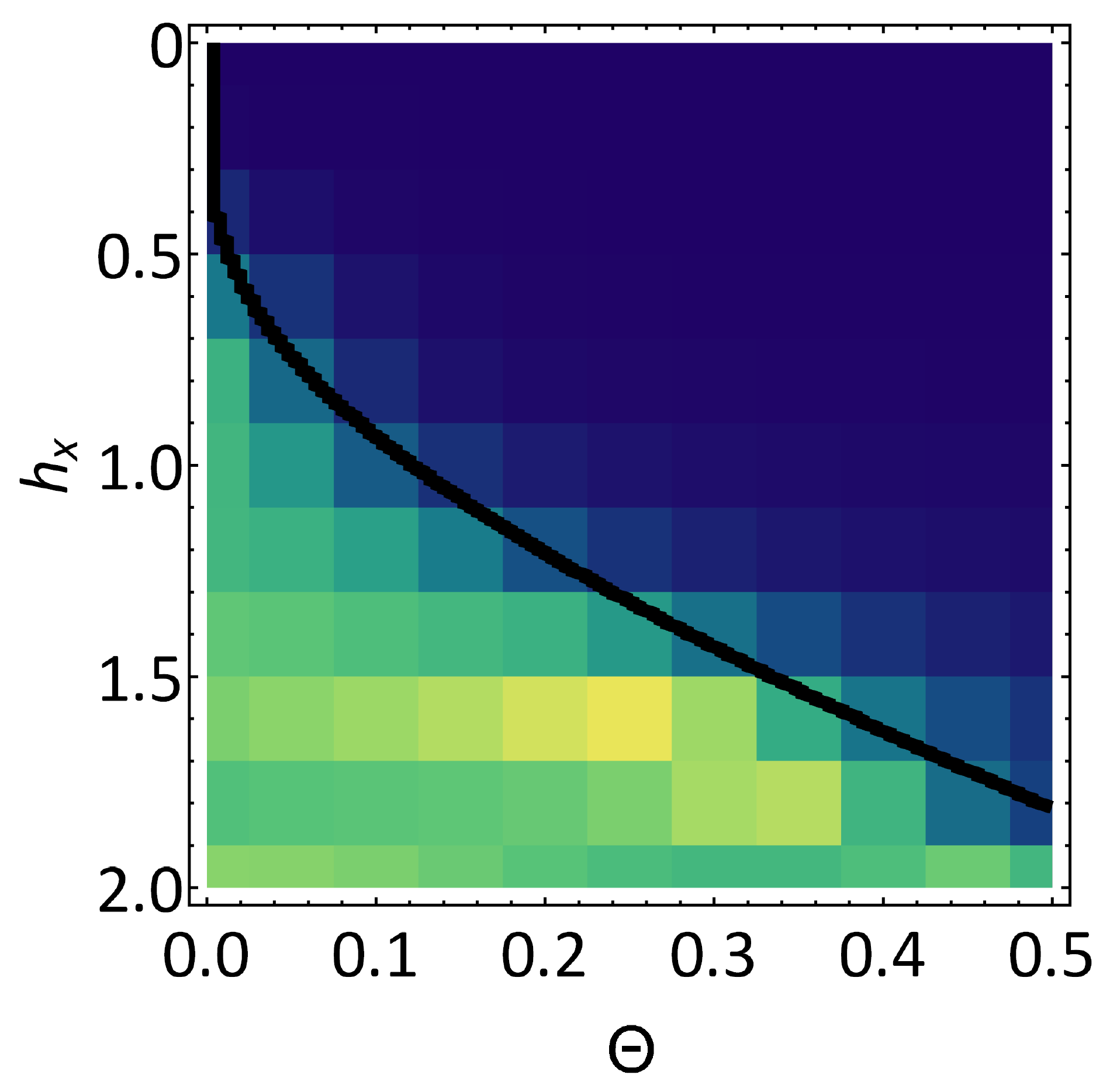}
\label{fig:S2_exact}
\end{minipage}
\hfill
\begin{minipage}[b]{.325\linewidth}
\centering\includegraphics[width=\textwidth]{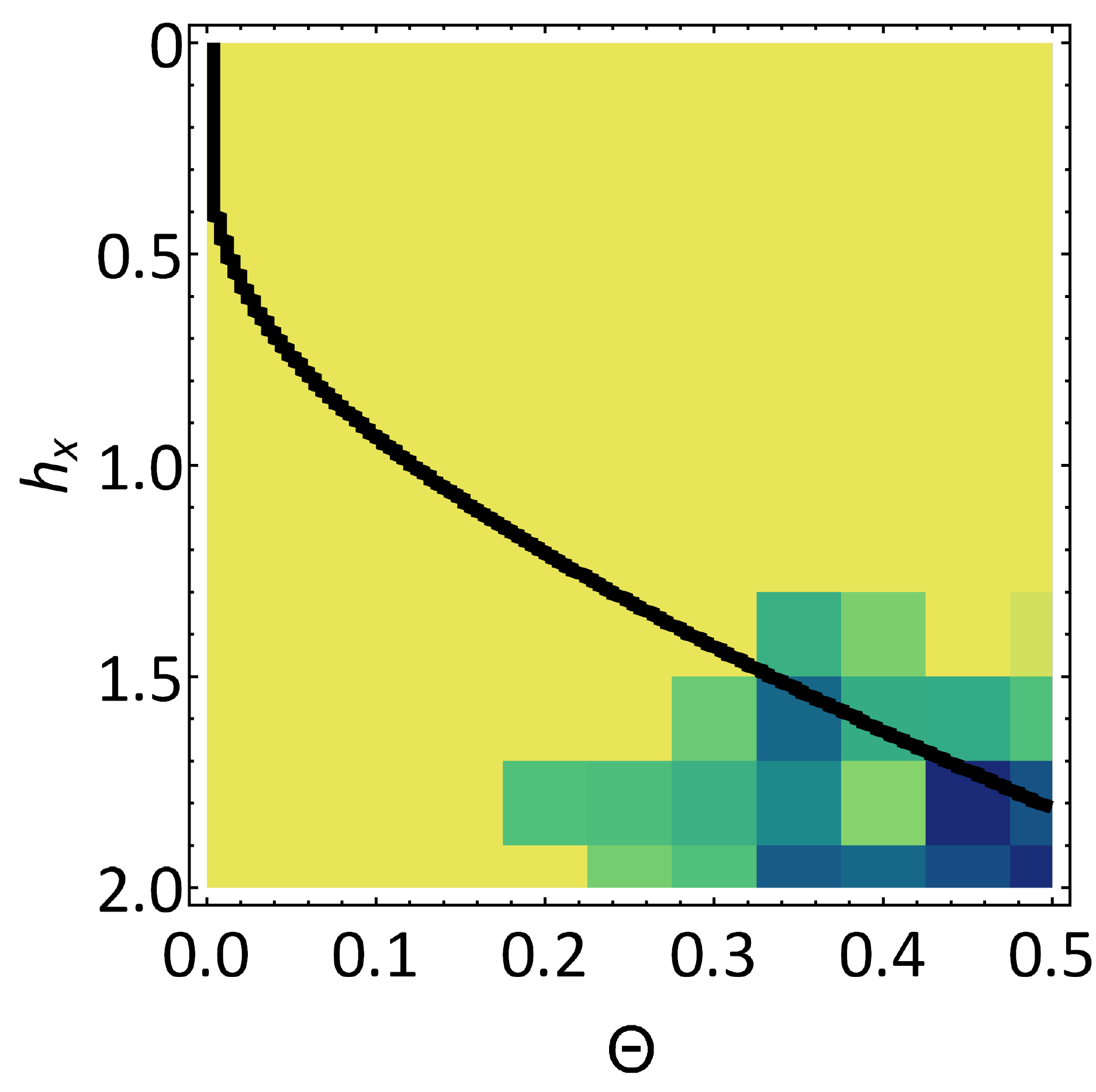}
\label{fig:Fidelity}
\end{minipage}
\hfill
\vspace{0.2cm}
\begin{minipage}[b]{.325\linewidth}
\centering\includegraphics[width=\textwidth]{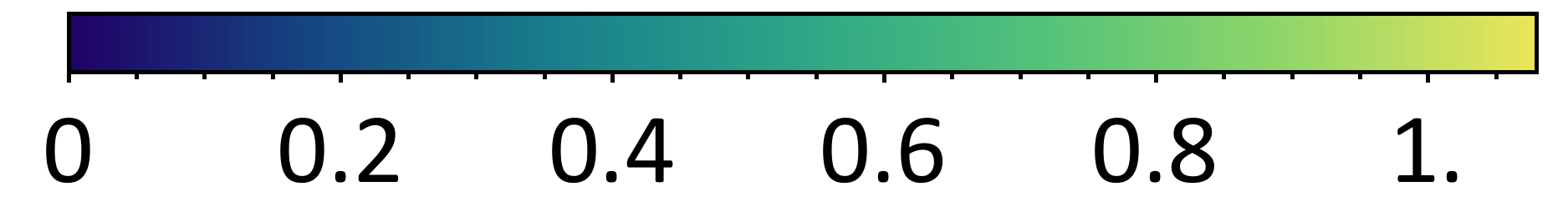}
\label{fig:Legend}
\end{minipage}
\caption{The second-order R\'{e}nyi entropy, $S_2$ calculated on a $N_{s} = 4$ lattice in the $h_{x}$-$\Theta$ plane.  Here the Trotter step size was $\delta t = 0.01$ and $S_{2}$ was measured on the 350th step. (left) The exact evolution; (middle) The calculation using the phase-damping method, selecting the best out of 350 runs for each point in parameter space;  (right) The fidelity between the exact evolution and the algorithm.}\label{fig:renyi2-compare}
\end{figure*}

In the two previously mentioned figures, the phase damping circuit was implemented with zero probability of measuring the auxiliary qubits in the ``ruined'' state.  This is the ideal case.  However, it is important to see how the algorithm can perform in the realistic case when the $\ket{1}$ state for the auxiliary qubit is measured with a non-zero probability.  To make this comparison, we calculated the second-order R\'{e}nyi entropy,
\begin{align}
\label{eq:renyi2}
    S_{2}(t) = -\log(\Tr[\rho^{2}(t)]),
\end{align}
as a function of evolution time. Here $\rho(t)$ is the reduced density matrix for an even bipartite split of the system at time, $t$.  We calculate this quantity in the $h_{x}$-$\Theta$ plane.  In Fig.~\ref{fig:renyi2-compare} we see this calculation and a comparison between three things:  On the left, evolution of the system using the circuit in Fig.~\ref{fig:qo-circuit}, performing a measurement on the 350\textsuperscript{th} time step of size $\delta t=0.01$, and choosing the iteration with the least number of ``1" measurements on the ancilla for each data point, out of 350 runs;  in the middle, the calculation of $S_{2}$  using the exact non-unitary evolution in the $h_{x}$-$\Theta$ plane; on the right, the fidelity,
\begin{equation}\label{eq:Fidelity}
    F(\rho,\sigma)=\Tr\sqrt{\rho^{1/2}\sigma\rho^{1/2}},
\end{equation}
between the density matrices obtained using the damping algorithm, and the one from exact evolution.  The fidelity above is symmetric in $\rho$ and $\sigma$. The black line denotes the exceptional line where the ground state and first excited state merge.

We can see that the algorithm reproduces the features of the exact evolution very well when $\Theta$ is small generally, and when $h_{x} \lesssim \Theta$.
Overall, we see that the fidelity is good over a modest range of the couplings, even in the presence of many 1-measurements on the ancillary qubit.


\subsection{Quantum circuit - Random Walk through time}
\label{sec:random-walk-ising}
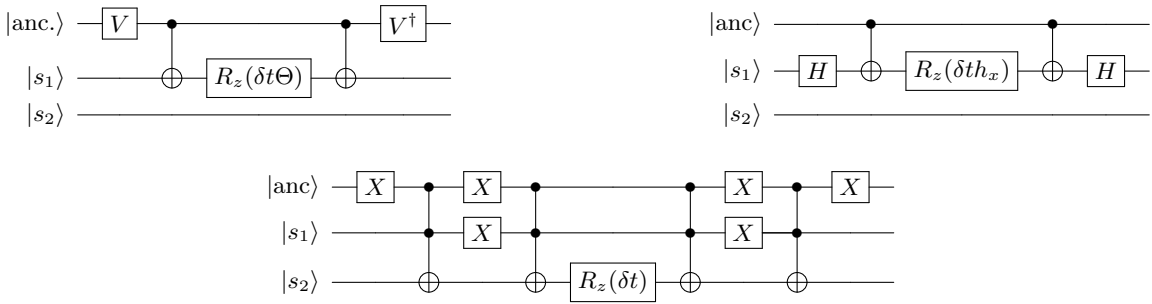
\begin{figure*}[thb]
\begin{minipage}[b]{.49\linewidth}
\centering
\begin{equation}
\nonumber
\Qcircuit @C=1em @R=.71em {
  \lstick{\ket{\text{anc.}}} & \gate{V} & \ctrl{1} & \qw & \ctrl{1} & \gate{V^{\dagger}} & \qw \\
  \lstick{\ket{s_{1}}} & \qw & \targ & \gate{R_{z}(\delta t \Theta)} & \targ & \qw & \qw \\
  \lstick{\ket{s_{2}}} & \qw & \qw & \qw & \qw & \qw & \qw 
}
\end{equation}
\label{fig:long}
\end{minipage}
\hfill
\begin{minipage}[b]{.49\linewidth}
\centering
\begin{equation}
    \nonumber
    \Qcircuit @C=1em @R=1em {
      \lstick{\ket{\text{anc}}} & \qw & \ctrl{1} & \qw & \ctrl{1} & \qw & \qw \\
      \lstick{\ket{s_{1}}}  & \gate{H} & \targ & \gate{R_{z}(\delta t h_{x})} & \targ & \gate{H} & \qw \\
      \lstick{\ket{s_{2}}} & \qw & \qw & \qw & \qw & \qw & \qw
    }
\end{equation}
\label{fig:trans}
\end{minipage}
\hfill
\begin{minipage}[b]{.5\linewidth}
\centering
\begin{equation}
\nonumber
    \Qcircuit @C=1em @R=.6em {
      \lstick{\ket{\text{anc}}} & \gate{X} & \ctrl{1} & \gate{X} & \ctrl{1} & \qw & \ctrl{1} & \gate{X} & \ctrl{1} & \gate{X} & \qw \\
      \lstick{\ket{s_{1}}} & \qw & \ctrl{1} & \gate{X} & \ctrl{1} & \qw & \ctrl{1} & \gate{X} & \ctrl{1} \qw & \qw & \qw \\
      \lstick{\ket{s_{2}}} & \qw & \targ & \qw & \targ & \gate{R_{z}(\delta t)} & \targ & \qw & \targ & \qw & \qw
    }
\end{equation}
\label{fig:three}
\end{minipage}%
\caption{Gate implementation for the random time walk Ising model in \EqnRef{eq:ising} with two system qubits and one ancillary qubit. $V$ diagonalizes the $\sigma_y$ Pauli matrix, and $H$ is the standard Hadamard gate.  The three circuits correspond, respectively, to the longitudinal field, the transverse field, and the nearest neighbor interactions.}\label{fig:IsingRW}
\end{figure*}

Here we discuss how to apply the random-walk algorithm to the transverse Ising model in an imaginary longitudinal field.
Using the procedure from Sec.~\ref{sec:rw-real}, we can expand the Hilbert space, and create a larger, Hermitian Hamiltonian from the non-Hermitian Hamiltonian in Eq.~\eqref{eq:ising}.
The new Hermitian Hamiltonian has three terms corresponding to a three-spin interaction and two, two-spin interactions.  The three-body interaction is the enlarged nearest-neighbor interaction in the original Ising model with the new ancillary qubit,
\begin{align}\label{eq:IsingNN}
    \hat{\ham}_{\text{N.N}} = -\hat{\sigma}^{z}_{\text{anc}} \sum_{\langle i j \rangle} \hat{\sigma}^{z}_{i} \hat{\sigma}^{z}_{j}.
\end{align}
An example of this interaction in a circuit is shown in the third image in Fig.~\ref{fig:IsingRW}.

The second term is a $Z$-$X$ interaction which comes from the transverse field in the original Ising model.  This interaction is between the ancillary qubit and a spin qubit,
\begin{align}
    \hat{\ham}_{\text{T}} = -h_{x} \hat{\sigma}^{z}_{\text{anc}} \sum_{i} \hat{\sigma}^{x}_{i}.
\end{align}
An excerpt of a circuit showing this part of the Hamiltonian can be seen in the second image in Fig.~\ref{fig:IsingRW}.
The final term is very similar, a $Y$-$Z$ spin-spin interaction coming from the longitudinal field,
\begin{align}
    \hat{\ham}_{\text{L}} = \Theta \hat{\sigma}^{y}_{\text{anc}} \sum_{i} \hat{\sigma}^{z}_{i}.
\end{align}
A figure showing the quantum circuit implementation of this interaction can be seen in the first image in Fig.~\ref{fig:IsingRW}.

Using the above circuits, we can simulate (in an error-free way) the random walk algorithm on a classical computer.  This allows us to assess the effect of the Trotter decomposition, and the repeated retracing of the system steps as it moves randomly forward and backward in time.  In practice, to simulate the random outcomes of measurement on the auxiliary qubit, we compute the reduced density matrix for that single qubit by tracing out the actual system, and reading the probabilities for measuring zero or one.  We then project the system accordingly into one of the two states, re-prepare the auxiliary qubit, and repeat the procedure.

\begin{figure}[t!]
    \centering
    \includegraphics[width=8.4cm]{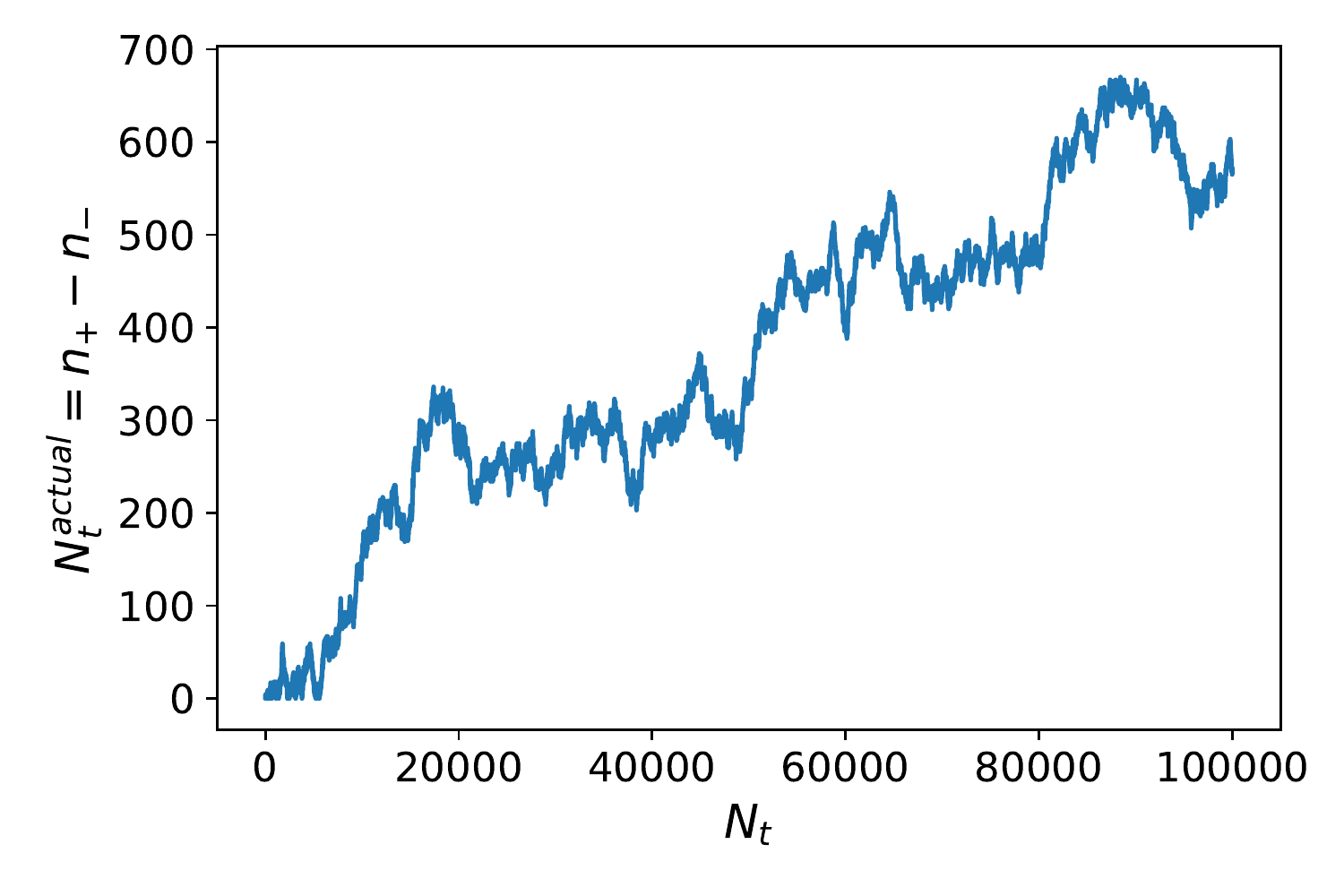}
    \caption{An example of how the system evolves in physical time, versus the number of Trotter steps actually taken in the computation.  On the $y$-axis we plot the difference between the number of forward steps and the number of backward steps, while on the $x$-axis show the number of Trotter steps taken.  This evolution is for a four-spin system with $h_{x} = 1.5$, $\Theta = 0.5$, and $\delta t = 0.001$.}
    \label{fig:act-steps}
\end{figure}

\begin{figure}[t!]
    \centering
    \includegraphics[width=8.6cm]{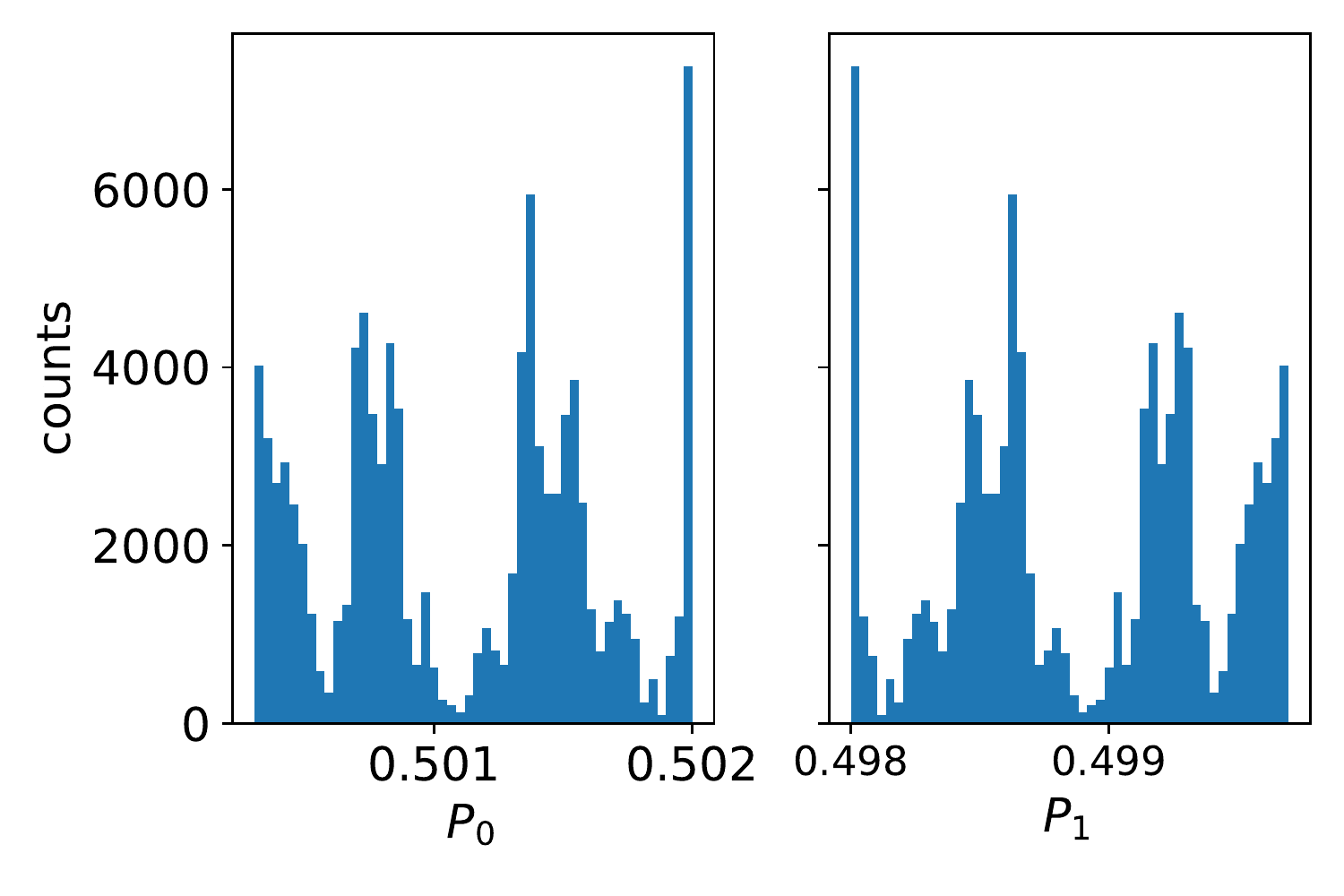}
    \caption{Histograms showing the probabilities for measuring zero or one over an example run of 100,000 steps.  We see a slight bias towards measuring the zero state, however this is only at $\mathcal{O}(\delta t)$.  These probabilities are for a four-spin system with $h_{x} = 1.5$, $\Theta = 0.5$, and $\delta t = 0.001$.}
    \label{fig:01probs}
\end{figure}

In Fig.~\ref{fig:act-steps} we show an example of typical evolution for the system.
In this case we have placed a ``mirror'' at $N_{t}^{\text{actual}}=0$, such that if the system would evolve into negative times we simply re-prepare the entire system in the initial state and begin again. On the $y$-axis is the number of physical times steps taken, while the $x$-axis is the number of actual Trotter steps taken in the computation.  We see it takes a great many steps to to move significantly forward in physical time.  Of course this is clear from the very nature of the random-walk algorithm,
and the probabilities can be seen in Fig.~\ref{fig:01probs}.
It's clear while there is an inherent asymmetry in the probabilities, they are approximately 50-50\% up to $\mathcal{O}(\delta t)$.  

\begin{figure}[t!]
    \centering
    \includegraphics[width=8.6cm]{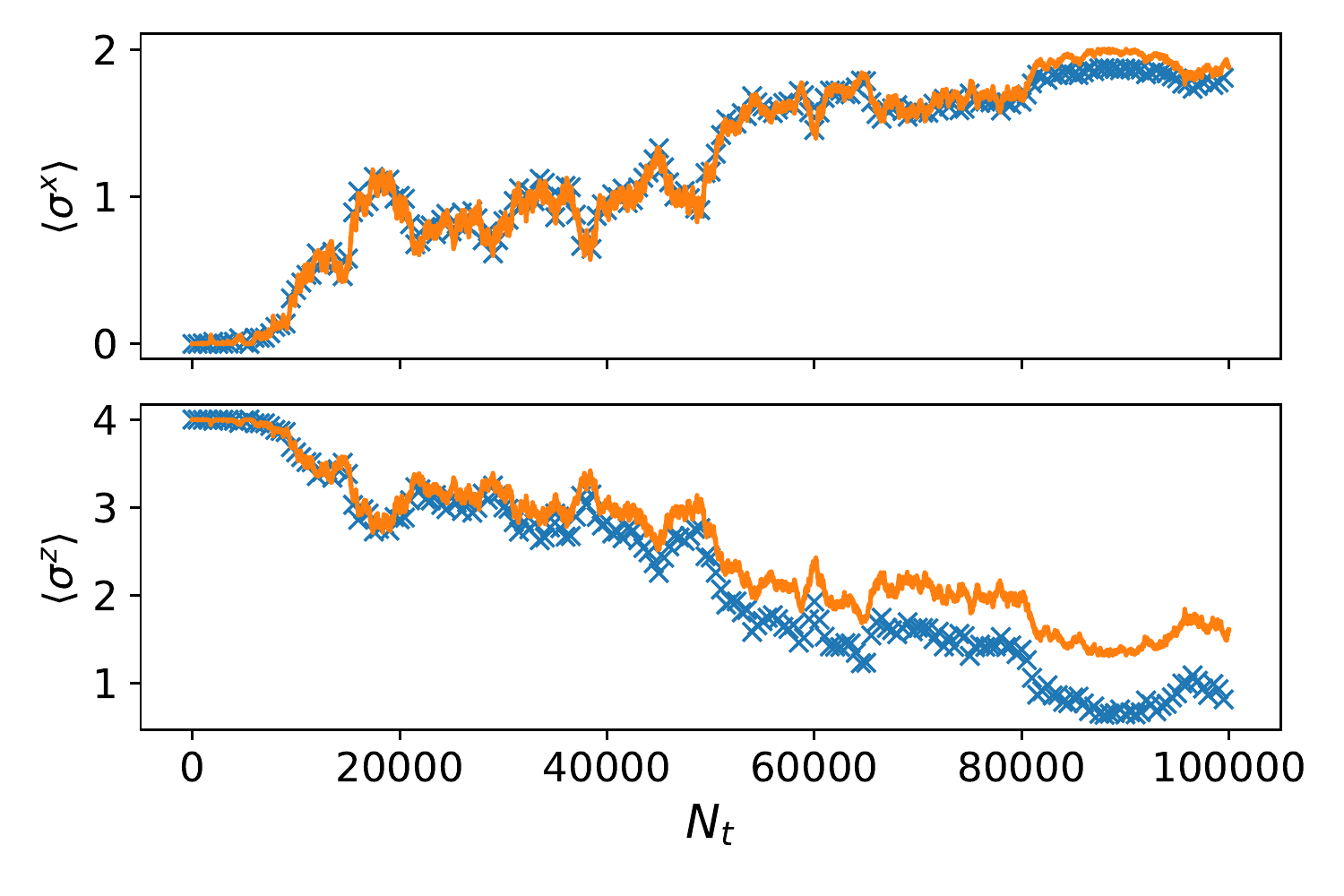}
    \caption{A comparison between the (error free) measurement of the average spin in the $x$ and $z$ direction (blue crosses), and the exact expectation value of the same quantities sampled at the same physical time (orange line) (See Fig.~\ref{fig:act-steps}).  These measurements are for a four-spin system with $h_{x} = 1.5$, $\Theta = 0.5$, and $\delta t = 0.001$.}
    \label{fig:rando-avg-x}
\end{figure}

\begin{figure}[t!]
    \centering
    \includegraphics[width=8.6cm]{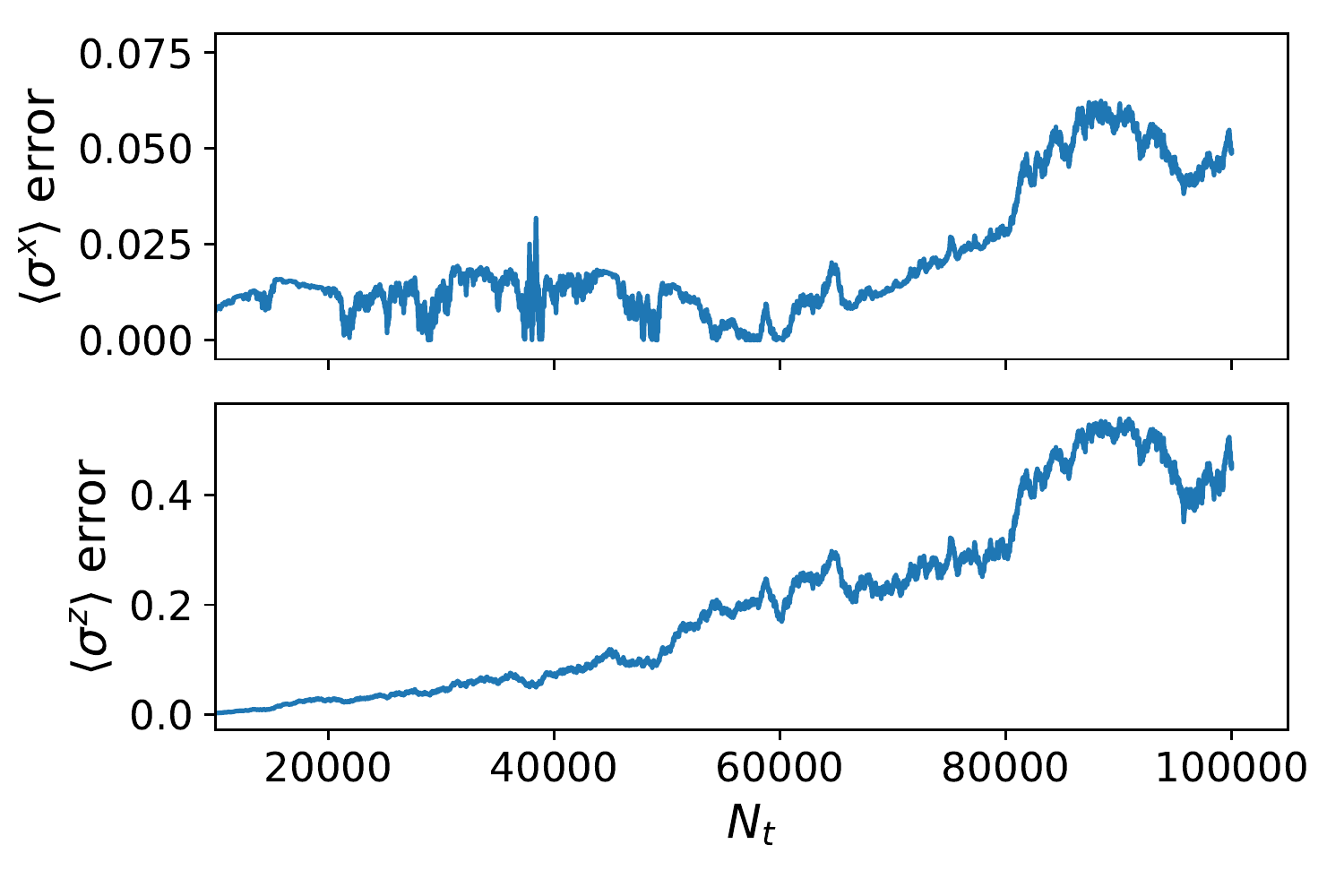}
    \caption{The error between the measured values and the exact values from Fig.~\ref{fig:rando-avg-x}.  We find reasonable quantitative agreement over a large number of Trotter steps; however, this is observable-dependent.}
    \label{fig:x-error}
\end{figure}

Nevertheless, the algorithm maintains quantitative agreement with the exact evolution.  An example of measured observables---the average spin along the $x$- and $z$-directions--can be seen in Fig.~\ref{fig:rando-avg-x}.
Here, we have plotted the error-free measurements (blue crosses) one can expect in the computation as a function of the actual number of Trotter steps that will be taken in the computation, along with the exact spin value (orange line) that moment in physical time (See Fig.~\ref{fig:act-steps}).  We plot the actual error associated with these observables in Fig.~\ref{fig:x-error}.


\section{Conclusion}
\label{sec:Conclusion}
We have presented three algorithms to simulate non-Hermitian Hamiltonian evolution on quantum computers using unital channels in conjunction with post-selection.  Both the System in Decline and the damping channel algorithms have the maximal approach in terms of probability of success of a single Trotter step to simulating the non-Hermitian Hamiltonian of interest.  The additional quantum jumps in these algorithms take one away from the desired evolution,
but if the imaginary coupling in the model is small or large relative to the real couplings, the approximate Hamiltonian that is simulated can possess similar characteristics.

For the random walk algorithm, each time step takes the system forward or backward in time according to the non-Hermitian Hamiltonian evolution.  
Because of the larger number of steps necessary to move forward in physical time, at least an order one factor of error is accumulated throughout the simulation when the number of times steps is $\gtrsim 1/\delta t$, and strict post-selection is necessary.  Nevertheless, for small physical times the results are in good agreement with exact calculations.

Using these algorithms we have studied a specific model, the one-dimensional Ising model with a real transverse field and a purely imaginary longitudinal field.  We found these algorithms are able to accurately reproduce global spin observables (\emph{e.g.} magnetization), as well as the second-order R\'{e}nyi entropy in the $h_{x}$-$\Theta$ plane where the Lee-Yang edge occurs.  We found the algorithms worked exceptionally well in the region of small non-Hermiticity where the imaginary coupling term is just a perturbation. We also found good agreement at large imaginary coupling relative to the real couplings, since the real exponential pulls the system back quickly to the desired ground state.  Finally, small physical times were simulated with excellent agreement since few errors have the oppurtunity to happen, and in that case the algorithm is exact up to Trotterization error. For larger values of the imaginary coupling where $\Theta \sim h_{x}$ (along with longer simulation times) a more extensive post-selection process is required.  These conditions make these algorithms excellent candidates for near-term quantum computing.

The above demonstrates that these algorithms can be used to simulate non-Hermitian systems on near-term devices, and in fact calculations are already underway for the Ising model studied here~\cite{wip}.  In addition, these algorithms allow for simulations in imaginary time (Euclidean time, or purely imaginary couplings) on quantum computing hardware.  Simulations at long Euclidean times force the system into its ground state, and so these algorithms could be useful for ground-state studies, or studies of slightly excited states.  Furthermore, the algorithms provide simple means to implement any non-unitary single- or two-qubit gate; however, the probability for success depends on the distance of the normalized eigenvalues from unity which generally could be quite large.


\section*{Acknowledgements} \label{sec:acknowledgements}
We would like to thank Erik Gustafson and Mike Hite for stimulating discussion while developing this work. We especially thank Hari Krovi for extensive discussion throughout the development of this project.  We thank Yannick Meurice, Erik Gustafson, Roni Harnik, and Simon Catterall for useful comments on the manuscript.   Jay Hubisz and Bharath Sambasivam are supported in part by U.S. Department of Energy (DOE), Office of Science, Office of High Energy Physics, under Award Number DE-SC0009998.  JUY was supported by the U.S. Department of Energy grant DE-SC0019139, and by Fermi Research Alliance, LLC under Contract No. DE-AC02-07CH11359 with the U.S. Department of Energy, Office of Science, Office of High Energy Physics.

\bibliography{Bibliography}
\end{document}